\documentclass[twocolumn,twocolappendix]{aastex631}
\usepackage{amsmath}
\usepackage{unicode-math}
\usepackage{dsfont}
\usepackage{relsize}
\usepackage{makecell}
\usepackage{multirow}

\begin{document}

\title{Role of magnetic reconnection in blazar variability using numerical simulation}

\author[0009-0006-6771-3365]{Chandan Kumar Das}
\affiliation{Indian Institute of Technology Indore, Khandwa Road, Simrol, Madhya Pradesh, India, 453552}

\author[0000-0001-5424-0059]{Bhargav Vaidya}
\affiliation{Indian Institute of Technology Indore, Khandwa Road, Simrol, Madhya Pradesh, India, 453552}

\author[0000-0002-5656-2657]{Amit Shukla}
\affiliation{Indian Institute of Technology Indore, Khandwa Road, Simrol, Madhya Pradesh, India, 453552}

\author[0000-0003-1454-6226]{Giancarlo Mattia}
\affiliation{Max-Planck Institute for Astronomy (MPIA), Königstuhl 17, 69117 Heidelberg, Germany}

\author[0000-0002-2950-6641]{Karl Mannheim}
\affiliation{Institute for Theoretical Physics and Astrophysics, Universität Würzburg, Campus Hubland Nord, Emil-Fischer-Str. 31, 97074 Würzburg, Germany}

\begin{abstract}
Fast $\gamma$-ray variability in blazars remains a central puzzle in high-energy astrophysics, challenging standard shock acceleration models. Blazars, a subclass of active galactic nuclei (AGN) with jets pointed close to our line of sight, offer a unique view into jet dynamics. Blazar $\gamma$-ray light curves exhibit rapid, high-amplitude flares that point to promising alternative dissipation mechanisms such as magnetic reconnection. This study uses three-dimensional relativistic magnetohydrodynamic (RMHD) and resistive relativistic magnetohydrodynamic (ResRMHD) simulations with the PLUTO code to explore magnetic reconnection in turbulent, magnetized plasma columns. Focusing on current-driven kink instabilities, we identify the formation of current sheets due to magnetic reconnection, leading to plasmoid formation. We develop a novel technique combining hierarchical structure analysis and reconnection diagnostics to identify reconnecting current sheets. A statistical analysis of their geometry and orientation reveals a smaller subset that aligns closely with the jet axis, consistent with the jet-in-jet model. These structures can generate relativistically moving plasmoids with significant Doppler boosting, offering a plausible mechanism for the fast flares superimposed on slowly varying blazar light curves. These findings provide new insights into the plasma dynamics of relativistic jets and strengthen the case for magnetic reconnection as a key mechanism in blazar $\gamma$-ray variability.

\end{abstract}


\section{Introduction} \label{sec:intro}

Astrophysical jets are supersonic, collimated outflows of plasma and magnetic fields originating near compact objects undergoing accretion. These jets are observed in different astrophysical sources like active galactic nuclei (AGNs), gamma-ray bursts (GRBs), microquasars (BH XRBs: black hole  X-ray binary systems) etc. The AGNs are the most  continuous astrophysical sources, emitting bolometric luminosities typically ranging from $10^{43}$ to $10^{48}$ erg/s \citep{Blandford2019, Kim2021}. Blazars are a class of AGNs with highly beamed relativistic jets oriented to the observer’s line of sight at a very small angle \citep{Urry1995}. The blazar emissions are strongly Doppler-boosted and predominantly non-thermal, encompassing the entire electromagnetic spectrum, ranging from low-energy radio frequencies to high-energy $\gamma$-rays. As the dominant part of the observed non-thermal emission from the AGN originates close to the central engine \citep{Sikora2007}, blazars offer unique opportunities to investigate the inner regions of AGNs and their jets.

Blazars are highly variable at all wavebands and at a wide range of time scales ranging from a few years \citep{Kniffen1993} to a few minutes \citep{Albert2007,Shukla2018}. Variability at all wavelengths is one of the defining properties of blazars. The discovery of very high energy (VHE; $> 100\,\rm{GeV}$) $\gamma$-ray flares from sources like PKS~2155-304 \citep{Aharonian2007}, Mrk 501 \citep{Albert2007} and high energy (HE; $0.1 - 300 \rm{GeV}$) $\gamma$-ray flares in sources like 3C 279 \citep{Ackermann2016}, CTA 102 \citep{Shukla2018} clearly highlights the extreme nature of these jets. These observational findings have motivated extensive studies to uncover the physical mechanisms responsible for blazars' rapid $\gamma$-ray variability. Additionally, the recent simultaneous observation of an IceCube neutrino event and $\gamma$-ray flare in TXS 0506+056 \citep{Aartsen2018} highlights the potential of blazars as multi-messenger sources \citep{Mannheim1993}.

The broad electromagnetic spectrum and diversity (different mass of central engine, redshift, etc.) of these objects can be explained by various models, including  shock acceleration \citep{Blandford1987}, star-in-jet model \citep{Barkov2012}, and mini jet-in-jet interaction from magnetic reconnection \citep{Giannios2009}. These fast variabilities observed in $\gamma$-rays are of particular significance as they manifest the fundamental physical processes governing the diverse temporal variability observed in blazars. Consequently, these also play a crucial role in constraining the mechanisms responsible for producing $\gamma$-rays within blazar jets. As observations suggest, many sources exhibit very short-duration flares, which indicate the size of compact emission or acceleration regions is too small compared to the size of the central black hole ($R_S = 2GM_{BH}/c^2$) \citep{Aleksic2011, Ackermann2016, Shukla2018, Agarwal2023}. Despite this, we consistently detect very high-energy $\gamma$-rays exceeding the threshold for electron-positron pair production, indicating that these high-energy photons can still survive within these compact emission regions. As we know, the optical depth to pair production ($\tau_{\gamma \gamma}$) is proportional to $L t_{var}^{-1} \Gamma^{-6}$ \citep{Begelman2008, Ackermann2016}, where $L$ is the observed $\gamma$-ray luminosity, $t_{var}$ is the characteristic variability timescale, and $\Gamma$ is the bulk Lorentz factor of the emitting region. Hence, for the survival of $\gamma$-rays, the Lorentz factors must be much larger than the typical values inferred from high-resolution radio observations ($\Gamma \sim 5$–$10$). In blazar jets, relativistic shocks with such high Lorentz factors are difficult to explain, posing severe constraints on the shock acceleration models to explain the relativistic particles. Plasmoids formed from magnetic reconnection filled with relativistic particles and magnetic fields can explain the extremely fast variability and help to alleviate the limitations faced by shock acceleration. This is also known as the mini jet-in-jet scenario driven by magnetic reconnection,  suggested to explain the puzzling short-time variability of gamma-rays observed with the Fermi large area telescope (\textit{Fermi}-LAT) \citep{Shukla2020}.

Relativistic magnetic reconnection, a process that involves the rapid reconstruction of magnetic field structure in plasma dominated by magnetic energy, has emerged as a widely applicable and promising mechanism for particle acceleration. In recent years, this phenomenon has been extensively studied using a variety of global numerical simulations \citep{Striani2016, BarniolDuran2017, Petropoulou2019, Bugli2025}. These investigations are primarily focused on determining the specific mechanisms that drive particle acceleration, such as the production of X- and O-points and the consequent release of energy when the astrophysical jets are highly turbulent due to current-driven kink instability \citep{Kadowaki2018, Bodo2020, Bodo2021, Kadowaki2021}.

\begin{figure}[ht!]
    \centering
    \includegraphics[width=1\linewidth]{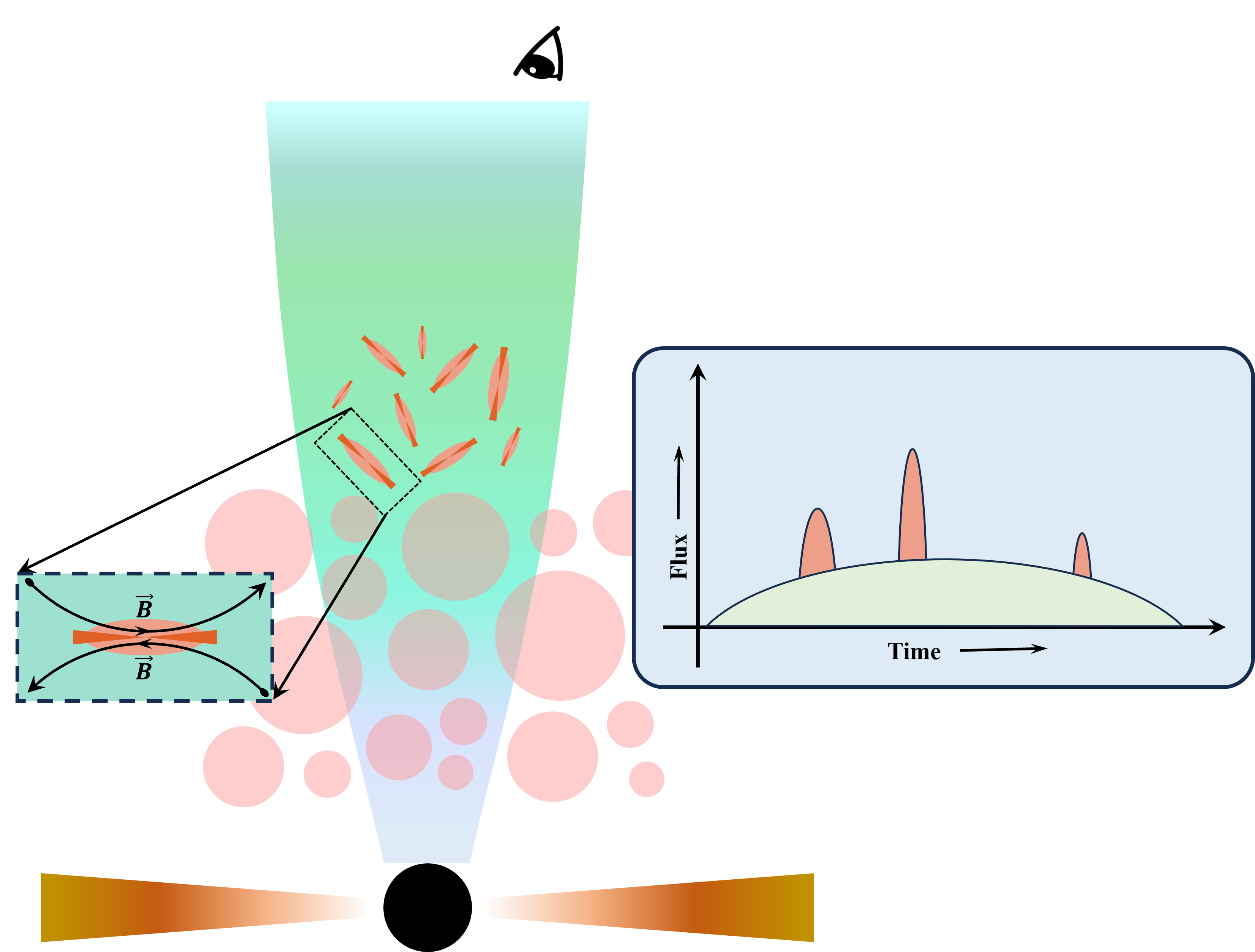}
        \caption{A schematic illustration of the inner region of a blazar jet and its associated light curve, showing the emission of an envelope flare produced by magnetic reconnection. The central black sphere represents the supermassive black hole. The orange shaded regions on either side depict the accretion disk, while the blue shaded structure, oriented perpendicular to the disk, represents the AGN jet. Surrounding the jet are several pink blobs that represent the broad-line region (BLR) clouds. Outside the BLR, small, randomly oriented orange structures indicate current sheets within the magnetic reconnection zone. Plasmoids can form within these current sheets, and when any of them move along the observer’s line of sight, they can produce very fast flares on top of the slowly varying envelope, as illustrated in the light curve shown on the right side box.}
    \label{fig:jet-in-jet model}
\end{figure}

Blazar jets remain highly collimated up to the broad line region (BLR) \citep{Blandford2019}; however, instabilities can develop outside the BLR, potentially leading to kink instabilities within the jet and the formation of filamentary current sheets \citep{Giannios2006, Giannios2013}. These current sheets (reconnection regions) move randomly with the direction of bulk flow of the jet, and within this reconnection region, magnetic energy heats the plasma and accelerates particles, leading to the formation of plasmoids. These plasmoids merge with each other to form a monster plasmoid, which then leaves the region along the current sheet with Alfv\'en speed ($V_A$) of the upstream flow. There is a recent work \citep{Kadowaki2021}, where they identified the reconnection region and calculated the Alfv\'en speed of these regions. Due to the random orientation of current sheets with our line of sight, the light curves show envelope flare structure and if the monster plasmoid is coming directly towards us, then there will be a fast flare on top of the envelope flare \citep{Giannios2009, Giannios2013, Narayan2012}. All these phenomena are well described as a pictorial form in figure \ref{fig:jet-in-jet model}.

Therefore, it is important to perform a detailed statistical analysis of geometric properties of current sheets formed from magnetic reconnection within the numerically simulated AGN jets. Although the resistivity within jets is very small (collisionless plasmas), this significantly affects the dynamics (turbulence) of the jets as the magnetic reconnection is directly influenced by the resistivity. In this regard, we have simulated a section of relativistic AGN jets (plasma column) using 3D relativistic magnetohydrodynamics (RMHD - only numerical resistivity) and resistive relativistic magnetohydrodynamics (ResRMHD) simulation. Several studies in the literature have focused on identifying current sheets in simulation data, primarily using current density profiles. A few have explored magnetic field topologies to locate potential reconnection regions. In this work, we adopt a novel approach combining current density analysis and reconnection diagnostics to robustly identify and characterize current sheets formed via magnetic reconnection in turbulent relativistic jets. These current sheets are particularly interesting as they are key sites for plasmoid formation and particle acceleration, potentially explaining the rapid gamma-ray variability observed in blazars.

The rest of the paper is structured as follows. Section~\ref{sec:Simulation} describes the numerical methods and setup used for the simulation of relativistic turbulent jets. Section~\ref{sec:Identification} outlines the methodology used to identify the current sheets responsible for magnetic reconnection, including a test case involving the Harris current sheet to validate the technique. In Section~\ref{subsec:Dynamics}, we present the dynamical results, including the evolution of jet energies, the development of kink instabilities, and the emergence of current sheets. Section~\ref{subsec:CurrentSheetDist} details the identification and distribution of current sheets across different simulations. Section~\ref{subsec:GeometricalDistribution} discusses the detailed statistics of the geometrical properties of the sheets, highlighting their potential implications for the observed fast variability in blazars. Finally, Section~\ref{sec:Discussion} summarizes the broader implications of our findings and concludes the study.

\section{Numerical simulation} \label{sec:Simulation}
To investigate current sheets formed from magnetic reconnection in relativistic turbulent jets, we perform three-dimensional simulations of a relativistic cylindrical jet using the {\tt \string PLUTO} code. This section outlines the numerical methods and setup, including variations in key input parameters. We also discuss the results from different simulation runs, focusing on the evolution of jet energies, the development of kink instabilities, and the emergence of reconnection-driven structures.

\subsection{Equations and numerical method}
We employ both RMHD and ResRMHD modules \citep{Mignone2019,Mattia2023} in the {\tt \string PLUTO} code\citep{Mignone2007} to solve the set of ResRMHD equations, facilitating the simulation of a segment of a relativistic jet (i.e., a relativistic plasma column). These equations are:

\begin{equation} \label{eq1}
    \begin{split}
    \frac{\partial D}{\partial t} + \nabla . (D \textbf{v}) &= 0,\\
    \frac{\partial\textbf{m}}{\partial t} + \nabla \cdot (w\textbf{uu} + pI + T) &= 0,\\
    \frac{\partial \mathcal{E}}{\partial t} + \nabla \cdot \textbf{m} &= 0,\\
    \frac{\partial \textbf{B}}{\partial t} + \nabla \times \textbf{E} &= 0,\\
    \frac{\partial \textbf{E}}{\partial t} - \nabla \times \textbf{B} &= -\textbf{J}.
    \end{split}
\end{equation}

Here, fluid conserved variables are the laboratory density $D = \rho\gamma$, total momentum density $\textbf{m} = w\gamma\textbf{u} + \textbf{E} \times \textbf{B}$, and total energy density, $\mathcal{E} = w\gamma^2 - p + u_{em}$. Again, $w$ is the specific enthalpy, $u_{em} = (E^2 + B^2)/2$ is the electromagnetic energy density, $I$ is the unitary matrix, and $T = -\textbf{EE}-\textbf{BB}+\frac{1}{2}(E^2 + B^2)I$ is the Maxwell's stress tensor. Here, 
$\rho$ is the fluid density, $p$ is the pressure, $\gamma$ is the Lorentz factor, $\textbf{u}$ is the fluid velocity, and $\textbf{E}$, $\textbf{B}$ and $\textbf{J}$ are electric field, magnetic field and current density in the laboratory frame respectively. \textit{Equation of state (EoS):} In this work, we adopted the ideal EoS and hence the specific enthalpy $w =\rho + \frac{\Gamma}{\Gamma - 1}p$ with a constant adiabatic index $\Gamma = 5/3$.

All the above equations follow flat space-time, and the four-velocity, four-electric field, four-magnetic field, and four-current density are as follows:

\begin{equation} \label{}
    \begin{split}
    &u^{\mu} = (\sqrt{1+u^2}, \textbf{u}) \equiv (\gamma, \gamma \textbf{v}),\\
    &e^{\mu} = (e^0,\textbf{e}) \equiv (\textbf{u}.\textbf{E}, \gamma \textbf{E} + \textbf{u} \times \textbf{B}),\\
    &b^{\mu} = (b^0,\textbf{b}) \equiv (\textbf{u}.\textbf{B}, \gamma \textbf{B} - \textbf{u} \times \textbf{E}),\\
    &J^{\mu} = (q, \textbf{J}) \equiv (q, q\textbf{v} + \eta^{-1}[\gamma \textbf{E} + \textbf{u} \times \textbf{B} - (\textbf{E}.\textbf{u})\textbf{v}])
    \end{split}
\end{equation}

Here, $u$, $e$, $b$, and $J$ are the velocity, electric field, magnetic field, and current density in the comoving frame, respectively, while $\textbf{v}$ denotes the three-velocity in the lab frame, and $\eta$ is the fluid resistivity. According to Ohm’s law in relativistic MHD, the comoving electric field is proportional to the current density: $e^{\mu} = \eta J^{\mu}$. In the non-resistive case of ideal MHD condition ($\eta \rightarrow 0$), this implies $e^{\mu} \rightarrow 0$, leading to the condition $\textbf{E} = -\textbf{v} \times \textbf{B}$. However, a finite $\eta$ yields a non-zero $e^{\mu}$ in the resistive case, thereby breaking this equality. Additionally, for post-processing, we use the Python-based module {\tt pyPLUTO} \citep{Mattia2025} to analyze the output data from {\tt PLUTO} simulations.

\subsection{Numerical setup : Initial and Boundary Conditions}
AGN jets exhibit high turbulence beyond the BLR, primarily due to the emergence of current-driven kink (CDK) instability. Therefore, our current study focuses on analyzing a 3D plasma column guided by a helical magnetic field, which is prone to instability induced by the current-driven kink mode. Motivated by insights from numerical simulations \citep{Striani2016,Bromberg2019,Bodo2019,Bodo2020,Bodo2021}, these configurations are likely to generate regions of magnetic reconnection capable of particle acceleration naturally. Our investigation begins with a fully rotating plasma column threaded by both poloidal ($B_z$) and toroidal ($B_{\phi}$) magnetic fields, with no radial magnetic field ($B_r = 0$), initialized using a three-dimensional force-free configuration in cylindrical coordinates \citep{Bodo2019}, as described in Appendix~\ref{Appendix:InitialSetup}.

Our configuration is characterized by the absolute value of the pitch on the axis $P_c$, the Lorentz factor on the axis $\gamma_c$, and the ratio of matter to magnetic energy density $M_a^2$.

\begin{equation} \label{eq3}
    P_c \equiv \left|\frac{rB_z}{B_{\phi}}\right|_{r=0}, \qquad M_a^2 \equiv \frac{\rho\gamma_c^2}{\left<\textbf{B}^2\right>}
\end{equation}
where, $\left<\textbf{B}^2\right> = \int_{0}^{r_j} (B_z^2+B_{\phi}^2)r \,dr/\int_{0}^{r_j} r \,dr$ represents the radially averaged magnetic energy density, with the jet radius set to $r_j = 1$. The Lorentz factor only due to the z-component of velocity is taken as,
\begin{equation}
    \gamma_z(r) = 1 + \frac{\gamma_c - 1}{cosh(r/r_j)^6},
\end{equation}
as defined in Appendix~\ref{Appendix:InitialSetup}(\ref{eqA2})

We adopt a 3D axisymmetric configuration in Cartesian coordinates, where $x$ and $y$ range from $-12r_j$ to $12r_j$, and $z$ ranges from $0$ to $24r_j$. The computational domain is discretized into a grid of $600\times600\times600$ uniform grids. Initially, the plasma column is characterized by a jet density $\rho_j = 1$, a Lorentz factor $\gamma_c = 5$, and $10$, pitch parameter $P_c = 1$, and $M_a^2 = 1$. The boundary conditions are set to be periodic in the $z$-direction and outflow in the $x$ and $y$ directions. We conduct simulations for three distinct plasma columns, each with varying resistivity values ($\eta = 0, 10^{-3}, 10^{-4}$).

For the $\eta = 0$ cases, we use the RMHD module, while for $\eta \neq 0$ we use the ResRMHD module. In both cases, linear reconstruction and $2^{nd}$ order time integration using a Runge-Kutta algorithm are used. For RMHD simulations, we use the less diffusive HLLD Riemann solver, which can capture more turbulent jet structures. For ResRMHD simulation, the set of ResRMHD equations~\ref{eq1} is solved with strong stability preserving IMplicit-EXplicit (IMEX) schemes. For this simulation, the MHLLC modified Riemann solver is used. A constrained-transport divergence-cleaning method is applied to both magnetic and electric fields in all simulations.

A precession perturbation \citep{Rossi2008} is applied in radial velocity, which drives the instability and induces turbulence. The radial velocity profile is given by, 
\begin{equation} \label{eq4}
    v_r = \frac{A}{24} \sum_{m=0}^2 \sum_{l=1}^8 cos(m\phi + \omega_lt + b_l)
\end{equation}
where, the amplitude of perturbation $A = \frac{\sqrt{(1+\epsilon)^2-1}}{\gamma_b(1+\epsilon)}$, $\epsilon = 0.05$, $m (= 0, 1, 2)$ is the azimuthal wavenumbers corresponding to different modes of perturbation, $\omega_l$ is a set of eight high and low frequencies, and $b_l$ is a randomly chosen phase shift. Following \citet{Rossi2008}, the eight frequencies are taken as $\omega_l = (0.5, 1, 2, 3, 0.03, 0.06, 0.12, 0.25) \, c_s$, where $c_s$ is the sound speed in the medium. Here, $\gamma_b$ denotes the bulk Lorentz factor of the jet, which in our analysis is taken to be equal to the axial Lorentz factor, $\gamma_c$.

In our simulations, the jet radius ($r_j$) taken at the outer edge of the BLR, which defines the unit length ($L_0$).To calculate this value, we have chosen the size of the BLR as $0.1 \, \mathrm{pc}$ (typically ranges from $0.03 \, \mathrm{pc}$ to $0.3 \, \mathrm{pc}$ \citep{Liu2006, Ghisellini1996}), the mass of the central supermassive black hole (SMBH) as $10^9 \, \mathrm{M_{\odot}}$ \citep{Ghisellini2009}, and a jet opening angle of $5^\circ$. The radius of the jet at the BLR is determined to be $r_j \sim 0.01 \, \mathrm{pc}$ using these values. The unit time ($t_0$) is defined as $t_0 = L_0 / v_0$, and the unit velocity ($v_0$) is set to the speed of light. The unit density is chosen as $\rho_0 \simeq 0.01 \,\mathrm{amu} \approx 1.66 \times 10^{-26} \,\mathrm{g\,cm^{-3}}$, and the corresponding unit magnetic field is defined as $B_0 = \sqrt{4\pi \rho_0 v_0^2} \approx 13.7 \,\mathrm{mG}$.

\subsection{Parameter Study}

We have conducted multiple simulation runs as listed in Table \ref{table:simulation_run} with various different parameters to study their effect on magnetic reconnection and current sheet formation in the jet. Here \textit{Res0g5p1} is our reference run with explicit resistivity $\eta = 0$, Lorentz factor $\gamma_c = 5$, and pitch parameter $P_c = 1$. Then we ran different simulations with different explicit resistivities and a higher Lorentz factor.

\begin{deluxetable}{  c  c  c  c  }
\renewcommand{\arraystretch}{1.2}
\tablecaption{Different simulation run.\label{table:simulation_run}}
\tablehead{
\colhead{\textbf{Run Name}} & 
\colhead{\makecell{\textbf{Explicit} \\ \textbf{Resistivity ($\eta$)}}} & 
\colhead{\makecell{\textbf{Lorentz} \\ \textbf{factor ($\gamma_c$)}}} & 
\colhead{\makecell{\textbf{Pitch} \\ \textbf{($P_c$)}}}
}
\startdata
Res0g5p1 & 0 & 5 & 1 \\ \hline
Res1g5p1 & $10^{-3}$ & 5 & 1 \\ \hline
Res2g5p1 & $10^{-4}$ & 5 & 1 \\ \hline
Res0g10p1 & 0 & 10 & 1 \\
\enddata
\end{deluxetable}

\section{Identification of current sheets} \label{sec:Identification}
Current sheets are narrow, high-gradient regions where the magnetic field undergoes rapid directional changes, often associated with magnetic reconnection and energy dissipation in plasma environments. Identifying such structures in simulation data is crucial for understanding reconnection-driven processes in astrophysical systems and bridging theoretical models with observational signatures. This has been attempted several times with different kinds of approaches for different types of research goals, such as \cite{Zhdankin2013}, \cite{Kadowaki2021}, \cite{Nurisso2023}. To achieve the goal of this work, we approach it with a unique and novel current-sheet identification technique, which can identify those current sheets within the magnetic reconnection region and find their geometrical properties. We used two methods; one is {\tt \string Astrodendro}, which can identify filamentary current sheets, and {\tt \string LoRD toolkit}, which can verify whether these sheets are within the reconnection region.

\subsection{Astrodendro} \label{astro}
Astrodendro \citep{Robitaille2019} is a structure identification algorithm written in Python based on the dendrogram technique in observed or simulated data. A dendrogram is a hierarchical tree structure that shows how smaller features combine into larger ones as the intensity threshold decreases, revealing the nested organization of structures in the data. Originally developed for hierarchical structure identification in star formation studies by \citet{Houlahan1992} using two-dimensional data, similar methodologies have been adapted across various astrophysical domains. For the first time, Astrodendro was used to identify structures in the Galactic Center cloud, a region noted for its high star-formation potential \citep{Johnston2014}. For instance, \citet{Rosolowsky2008} applied analogous techniques to three-dimensional molecular-line position–position–velocity data cubes, while \citet{Piperno2020} utilized them in the study of giant molecular clouds. More recently, \citet{Paul2022} employed a similar dendrogram-based approach to identify three-dimensional volumetric structures of flux transfer events (FTEs) in global MHD simulations of planetary magnetospheres, demonstrating its applicability in plasma environments. In general, the algorithmic framework can theoretically be applied to any generalized input dataset.

The output of the algorithm is represented as a dendrogram, which abstracts the hierarchical structure in the input data into a structure tree. This structure tree consists of smaller entities referred to as \emph{branches}, which can be further decomposed into independent structures known as \emph{leaves}. In order to find these independent structures, we have to specify the minimum height and pixel number of the structure, i.e., the threshold value above which any patch of the data is considered an independent structure. In our study, we extend the application of this structure-finding algorithm to detect three-dimensional current sheets in a global MHD simulation of relativistic jets for the first time, to our knowledge. For this purpose, we utilize the astrodendro python package \footnote{The astrodendro package and its documentation can be found here: \url{https://dendrograms.readthedocs.io/en/stable/}} \citep{Robitaille2019}.

\subsection{LoRD toolkit}
LoRD (Locate Reconnection Distribution)\footnote{The LoRD toolkit and its documentation can be found here: \url{https://github.com/RainthunderWYL/LoRD}} \citep{Wang2024a}, is a Matlab toolkit available on GitHub designed for efficiently analyzing 3D magnetic reconnection based on the general reconnection theory proposed by \cite{Hesse1988}. The analyze reconnection distribution (ARD) function of this toolkit takes inputs of $B1$, $B2$, and $B3$ matrices representing discrete magnetic field data on three dimensions: meshgrid data. Additionally, it requires $X1$, $X2$, and $X3$, which are 1D coordinate arrays along the three directions. Notably, LoRD employs an efficient numerical method solely involving algebraic manipulations on the discrete magnetic field, thereby avoiding computationally expensive operations such as field-line tracing and root-finding. As a result, this method directly provides the classification and geometric parameters of local magnetic fields at arbitrary grid points while accurately locating reconnection sites.


\subsection{Test Case for Identifying Current Sheets} \label{Ident_currentsheet}

To test our identification technique, we perform a pseudo-3D simulation (i.e., a 2D simulation extended into the third dimension with a small thickness) of the Harris current sheet in the ideal MHD regime (no explicit resistivity). The initial conditions are adopted from \citet{Puzzoni2021}. The simulation domain spans a rectangular box of size $L \times L/2 \times L/8$ along the $x$-, $y$-, and $z$-axes, respectively, with a resolution of $1024 \times 512 \times 128$ grid points. Here, the $x$–$y$ plane captures the primary two-dimensional structure of the current sheet. At the same time, the $z$-direction represents a narrow but resolved third dimension to allow for the identification of three-dimensional features. The equilibrium magnetic field follows a Harris-sheet profile: $B_x(y) = B_c tanh\left(\frac{y}{a}\right)$. Here, $a$ is the initial width of the current sheet and $B_c$ is the initial normalized magnetic field strength. Similarly, the pressure term is denoted as $P(y) = \frac{1}{2}B_c^2(\beta + 1) - \frac{1}{2}B_x^2(y)$, which maintains constant pressure through the sheet; here, $\beta$ is taken as $0.01$. The boundary conditions in the x-direction and y-direction are periodic and reflective, respectively.

The current density profile is crucial for understanding the jet's magnetic structure and dissipation processes. To explore the formation and structure of current sheets in our simulations, we analyze the spatial distribution of the current density $\mathbf{J}$, which serves as a direct tracer of magnetic field gradients and reconnection-prone locations. In the ideal MHD limit, the current density is calculated using Ampère's law as follows:

\begin{equation} \label{CD}
    \mathbf{J} = \nabla \times \mathbf{B},
\end{equation}

where $\mathbf{B}$ represents the magnetic field vector. Numerically, the curl is calculated using central finite differences across the computing grid. To capture localized enhancement in current, we evaluate the magnitude of the current density, $|\mathbf{J}| = \sqrt{J_x^2 + J_y^2 + J_z^2}$, which serves as an indicator of regions where magnetic gradients steepen and dissipation is expected. High values of $|\mathbf{J}|$ imply strong magnetic field gradients and are commonly associated with the formation of thin current sheets, confined structures where magnetic reconnection may become efficient.

We begin by computing the current density profile from the simulation data to identify current sheets associated with magnetic reconnection. The {\tt Astrodendro} algorithm is then applied with a specified minimum current density threshold to extract distinct structures, interpreted as current sheets. This set is hereafter referred to as $S_{\rm Ad}$. To assess whether these sheets correspond to active reconnection sites, we use the {\tt LoRD toolkit}, which identifies localized reconnection regions, denoted as $S_{\rm L}$. For further analysis, we consider only those current sheets that spatially overlap with reconnection sites, i.e., $S_{\rm Ad} \, \cap \, S_{\rm L}$. The current density threshold in {\tt Astrodendro} is chosen such that at least $85 \%$ of the identified current sheets overlap with reconnection regions, ensuring the reliability of the selected structures.

\begin{figure}[ht!]
    \centering
    \includegraphics[width=1\linewidth]{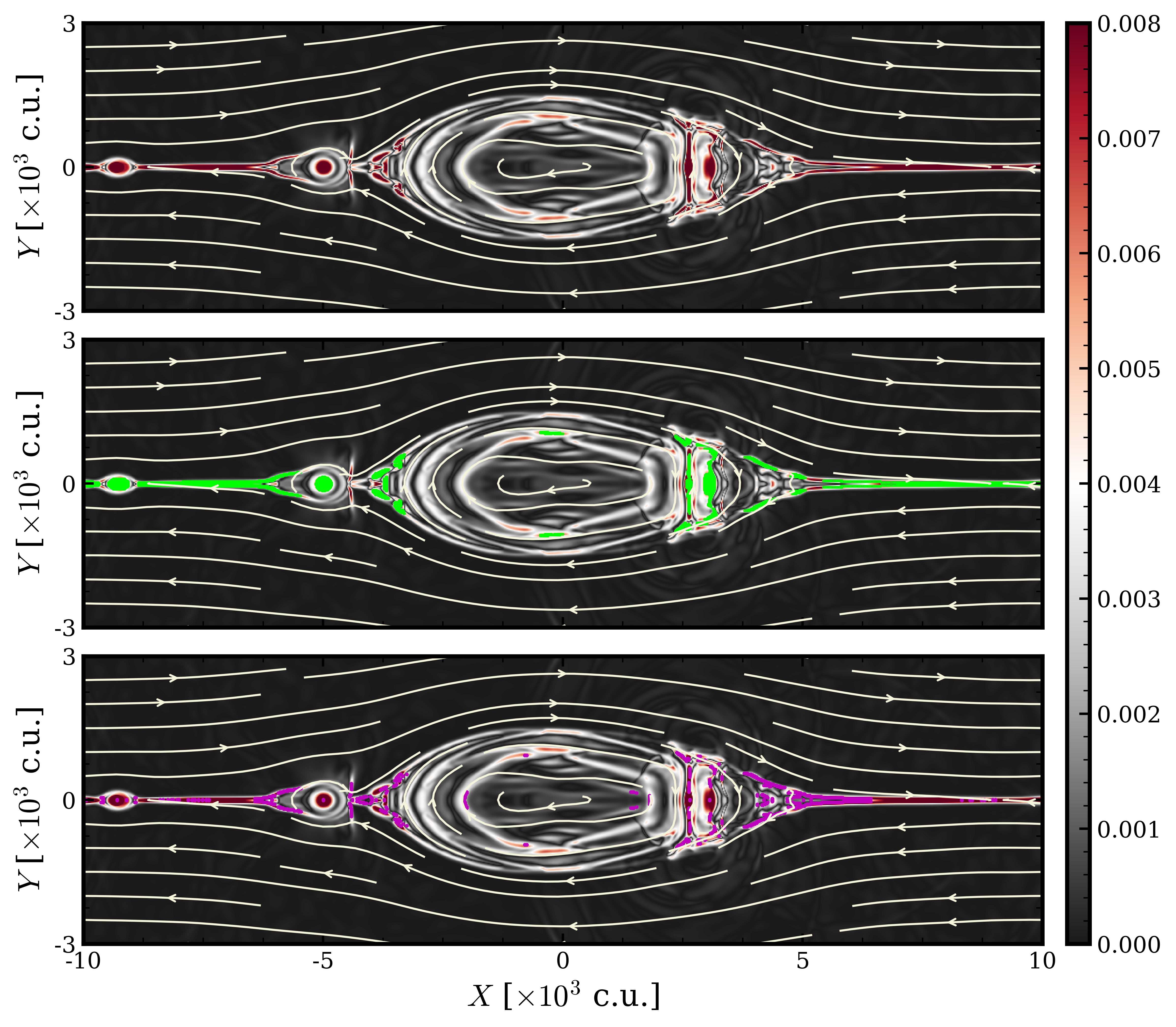}
    \caption{2D slice from the Harris current sheet simulation at $t = 1.4 \times 10^5$. The background colormap shows current density profile; beige arrows indicate magnetic field lines. Middle panel: green regions denote current sheets identified by {\tt Astrodendro}. Bottom panel: magenta points mark reconnection sites identified by the {\tt LoRD toolkit}.}
    \label{fig:2d_reconnection}
\end{figure}

Figure~\ref{fig:2d_reconnection} displays three subplots at simulation time $t = 1.4 \times 10^5$ (in code units), when the current sheet is undergoing fast magnetic reconnection, characterized by the formation of multiple plasmoids. The background color map in all panels illustrates the current density profile, with the colorbar denoting its strength in code units. Beige streamlines with arrows depict the magnetic field lines and their directions. The middle panel features green structures representing current sheets identified using the {\tt Astrodendro} method. The magenta points in the bottom panel indicate reconnection locations found by the {\tt LoRD toolkit}. The middle and bottom panels of Figure~\ref{fig:2d_reconnection} demonstrate that almost $90\%$ of the current sheets identified by {\tt Astrodendro} overlap with reconnection regions detected using the {\tt LoRD} toolkit. This result is consistent with the threshold condition defined for spatial overlaps between the two methods. Moreover, all identified current sheets also overlap with regions of enhanced current density, as shown in the top panel of Figure~\ref{fig:2d_reconnection}, indicating the reliability of both detection techniques.

\section{Results} \label{sec:results}

\subsection{Dynamics} \label{subsec:Dynamics}
\begin{figure*}[ht!]
    \centering
    \includegraphics[width=1\linewidth]{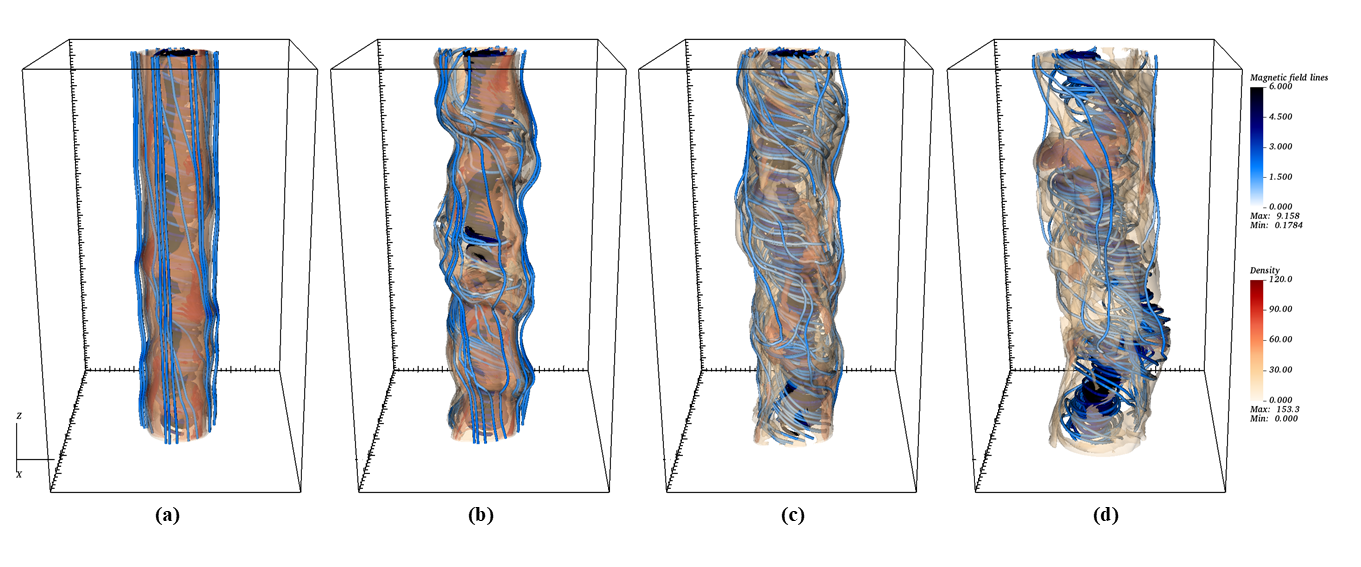}
    \caption{Three-dimensional visualization of magnetic field lines (colored using a blue-hot color scale in the unit of $B_0$) overlaid with isosurfaces of fluid density (orange-red color scale in the unit of $\rho_0$) at four different time steps for the simulation run \textbf{Res0g5p1}. The snapshots correspond to: (a) $t = 123 \, t_0$, (b) $t = 153 \, t_0$, (c) $t = 183 \, t_0$, and (d) $t = 213 \, t_0$. These time steps capture the evolution of the magnetic field and the development of kink instability within the jet.}
    \label{fig:Res0g5p1_MF_evolution}
\end{figure*}

We begin by discussing the results from the reference run \textit{Res0g5p1} (see Table \ref{table:simulation_run}), followed by a comparative analysis with the other simulation runs. Figure \ref{fig:Res0g5p1_MF_evolution} illustrates the time evolution of magnetic field lines and fluid density in the relativistic jet for the \textit{Res0g5p1} case, shown at four different simulation time steps: $t = 123$, $153$, $183$, and $213 \, t_0$. In figure \ref{fig:Res0g5p1_MF_evolution}, the magnetic field lines are displayed in blue-hot color scales, and the iso-surfaces of the fluid density in orange-red color scales. At early times, the magnetic field lines form a helical structure along the jet spine. As systems evolve with time, the CDK instability develops, deforming and distorting the helical configuration of magnetic field lines, resulting in the dissipation of electromagnetic (EM) energy into thermal and kinetic energy. Due to this instability, magnetic reconnection may occur within the jet.

\begin{figure}[ht!]
    \centering
    \includegraphics[width=1\linewidth]{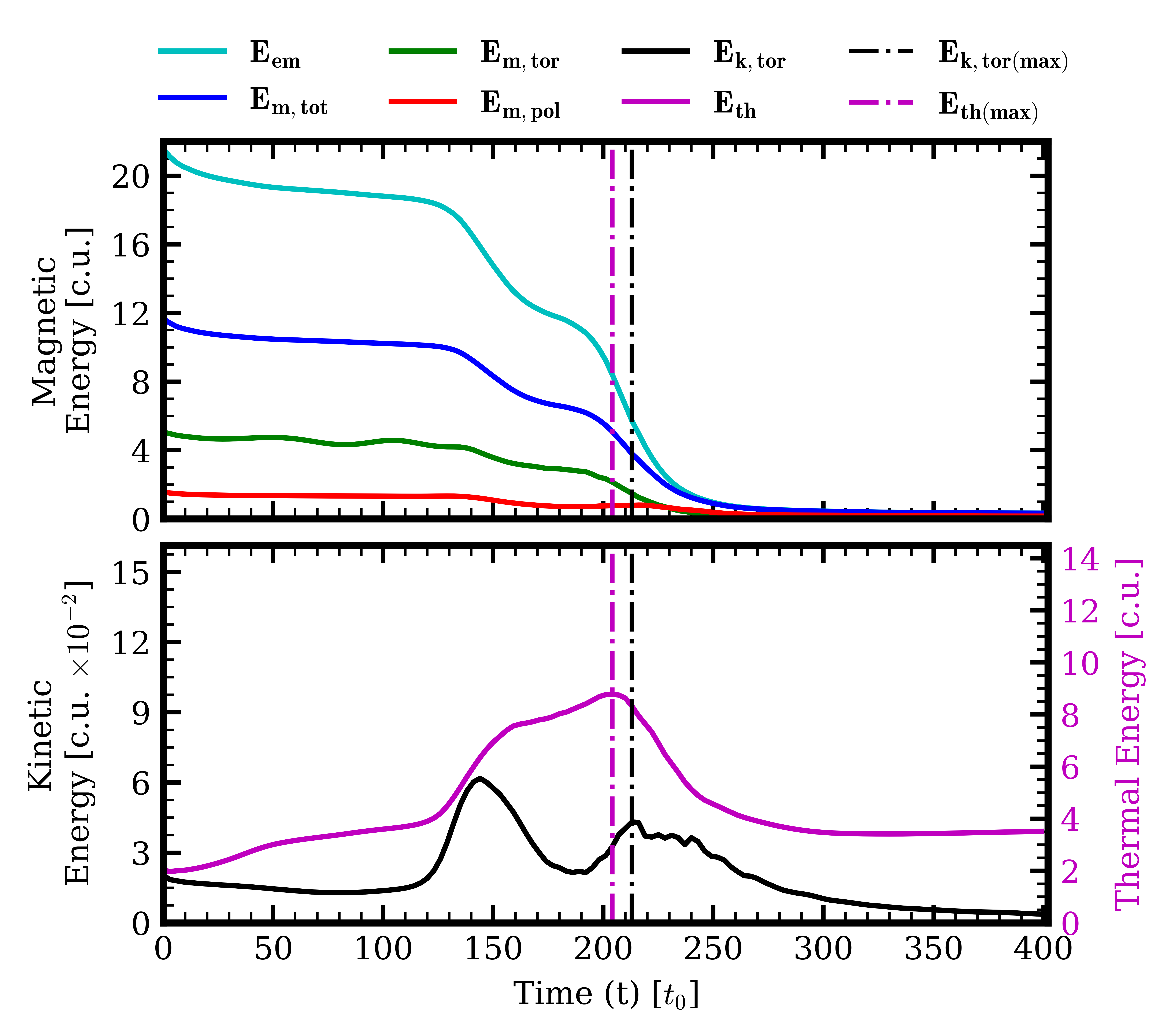}
    \caption{Time evolution of the different energy components within the AGN jet for the simulation run \textbf{Res0g5p1}. \textit{Top panel:} Evolution of magnetic energy components. \textit{Bottom panel:} Evolution of thermal energy and toroidal component of kinetic energy. The vertical maroon and black dash-dotted lines denote the epochs corresponding to the maximum of thermal energy and the subsequent local maximum of toroidal kinetic energy, respectively. All quantities are volume-averaged within the jet and normalized to code units.}
    \label{fig:Res0g5p1_E_evolution}
\end{figure}

To identify the onset and evolution of the CDK instability, we analyze the temporal evolution of various volume-averaged energy components within the jet. Figure~\ref{fig:Res0g5p1_E_evolution} presents the time evolution of these energy components, normalized to code units. The top panel shows the evolution of toroidal (solid green), poloidal (solid red), total (solid blue) magnetic energies, and the EM energy (solid cyan), defined as follows:

\begin{equation} \label{MagEnergy}
    \begin{split}
    E_{m,tor} = \frac{1}{V}\int_V \frac{B_{\phi}^2}{2}dxdydz, \\ E_{m,pol} = \frac{1}{V}\int_V \frac{B_{z}^2}{2}dxdydz,
    \\ E_{m,tot} = \sqrt{E_{m,tor}^2 + E_{m,pol}^2}
    \end{split}
\end{equation}
Where, $\textbf{B}$ is the magnetic field vector, $\textbf{u}$ is the velocity, $V$ is the total volume of the system, $B_{\phi}$ is the toroidal, and $B_z$ is the poloidal magnetic field. For the ideal RMHD case, the total electromagnetic energy is computed as:

\begin{equation} \label{EMEnergy1}
    E_{em} = \frac{1}{V}\int_V \frac{\textbf{B}^2 + [\textbf{u}^2 \textbf{B}^2 - (\textbf{u}.\textbf{B})^2]}{2}dxdydz,
\end{equation}
whereas for the resistive case, it is given by:

\begin{equation} \label{EMEnergy2}
    E_{em} = \frac{1}{V}\int_V \frac{\textbf{B}^2 + \textbf{E}^2}{2}dxdydz,
\end{equation}

Similarly, the bottom panel represents the time evolution of toroidal kinetic (solid black) and thermal energy (solid maroon), which are given by,
\begin{equation} \label{KinEnergy}
    E_{k,tor} = \frac{1}{V}\int_V \gamma_{\phi}(\gamma_{\phi}-1) \rho dxdydz,
\end{equation}
and,
\begin{equation} \label{ThEnergy}
    E_{th} = \frac{1}{V}\int_V [(h-1) \gamma^2 \rho]dxdydz,
\end{equation}

Where, $E_{k,tor}$ is the toroidal kinetic energy and $E_{th}$ is the thermal energy in the cold gas limit \citep{Mignone2007a, Dubey2023}, which depends upon the specific enthalpy, $h = 1 + \frac{\Gamma}{\Gamma - 1} \frac{p}{\rho}$ for the ideal gas equation of state (EOS). Here, $\Gamma (=5/3)$ is the gas constant, $p$ is the pressure, $\rho$ is the density, $\gamma_{\phi}$ is the toroidal component of the Lorentz factor, and $\gamma$ is the total Lorentz factor.

In the bottom panel of Figure~\ref{fig:Res0g5p1_E_evolution}, we observe a sudden rise in both the toroidal kinetic and thermal energy components beginning around $t = 123 \, t_0$, accompanied by a concurrent decline in magnetic and EM energies. Following this phase, the toroidal kinetic energy reaches a peak before gradually decreasing. Interestingly, despite these energetic changes, there is no prominent signature of kink instability at this time, as seen in panel (a) of Figure~\ref{fig:Res0g5p1_MF_evolution}, which depicts the magnetic field configuration at $t = 123 \, t_0$. A second phase of energetic activity is initiated around $t = 190 \, t_0$, during which both toroidal kinetic and thermal energies increase again. The thermal energy reaches its maximum at $t = 204 \, t_0$, marked by a vertical maroon dash-dot line. In contrast, the toroidal kinetic energy attains a local maximum at $t = 213 \, t_0$, denoted by a vertical black dash-dot line. This period also corresponds to a notable drop in magnetic and EM energies. During this phase, we clearly observe the development of kink instability, as shown in panel (d) of Figure~\ref{fig:Res0g5p1_MF_evolution}, where the helical magnetic field structure becomes significantly distorted near the jet spine. These observations suggest that magnetic reconnection is dominant during this later phase, likely triggered or enhanced by the development of kink instability. The simultaneous decrease in magnetic and EM energies and increase in kinetic and thermal energies further support this interpretation, indicating energy conversion and dissipation processes associated with reconnection and plasma heating.

\begin{figure*}[ht!]
    \centering
    \includegraphics[width=1\linewidth]{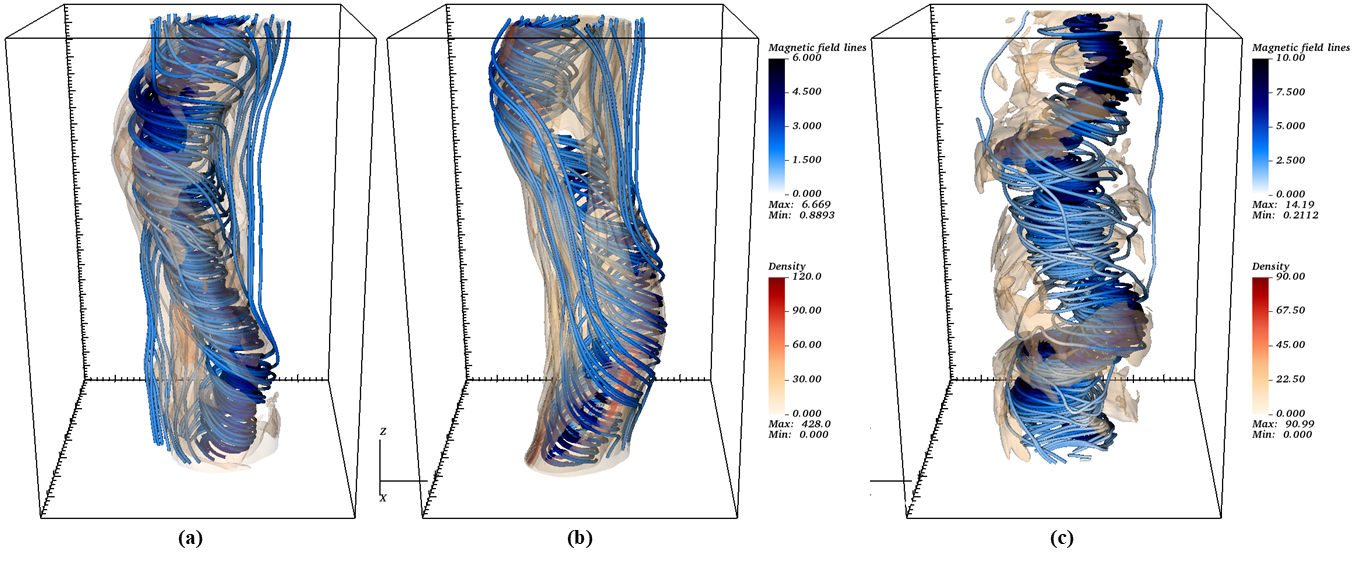}
    \caption{Three-dimensional representation of magnetic field lines (colored using a blue-hot color scale in the unit of $B_0$) and isosurfaces of fluid density (orange-red color scale in the unit of $\rho_0$) for three different simulation runs at the time when the kink instability is well developed within the jet. From left to right: \textbf{Res2g5p1}, \textbf{Res1g5p1}, and \textbf{Res0g10p1}. These visualizations highlight the structural deformation of the jet due to the onset of kink instability under varying resistivity and Lorentz factor conditions.}
    \label{fig:Res12_0g5_10p1_MFlines}
\end{figure*}

\begin{figure*}[ht!]
    \centering
    \includegraphics[width=1\linewidth]{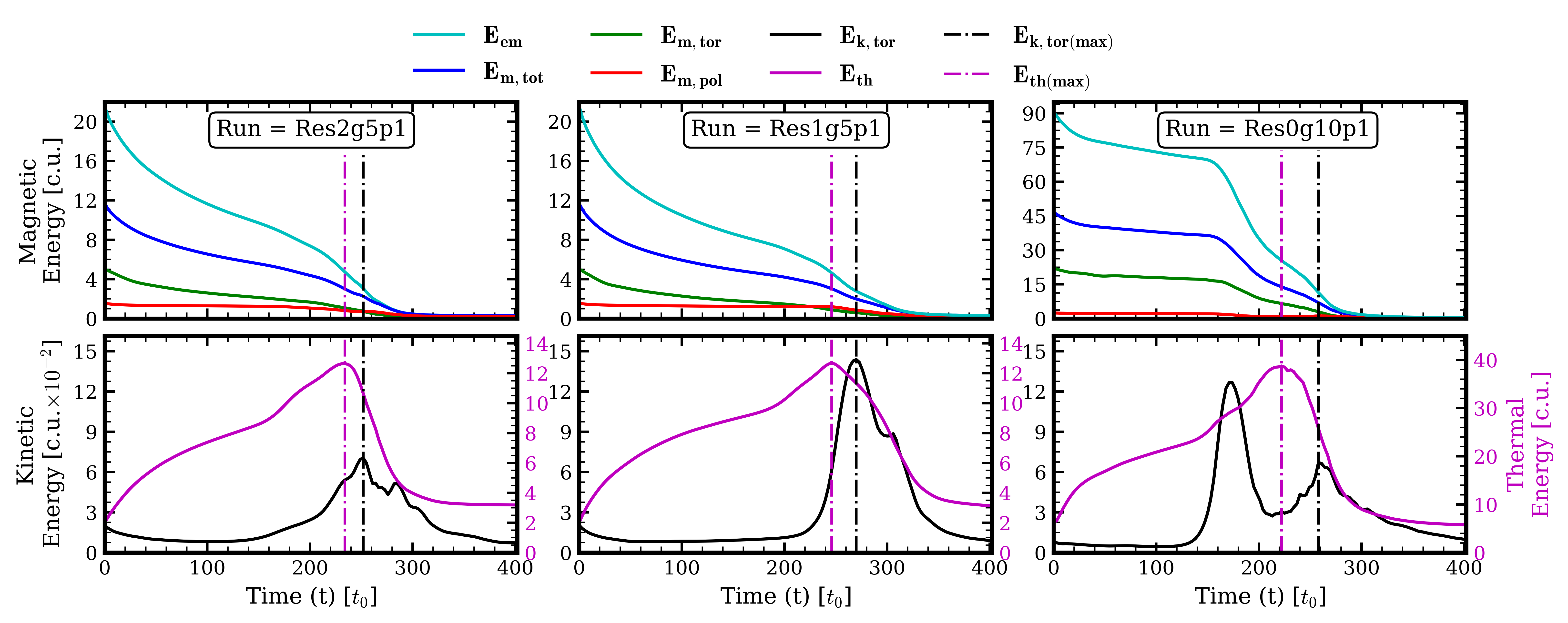}
    \caption{Time evolution of different energy components within the AGN jet for three distinct simulation runs. \textit{Top panels:} magnetic energies; \textit{bottom panels:} thermal energy and toroidal kinetic energy. The vertical maroon and black dash-dotted lines denote the epochs corresponding to the maximum of thermal energy and the subsequent local maximum of toroidal kinetic energy, respectively. From left to right, the columns correspond to simulation runs \textbf{Res1g5p1}, \textbf{Res2g5p1}, and \textbf{Res0g10p1}, respectively.}
    \label{fig:Res12_0g5_10p1_E_evolution}
\end{figure*}

We have also conducted three additional simulations by varying the jet's effective resistivity and axial Lorentz factor. In two of these simulations, \textit{Res2g5p1} and \textit{Res1g5p1} (listed in Table~\ref{table:simulation_run}), we incorporated explicit resistivity values of $10^{-4}$ and $10^{-3}$, respectively, while keeping the axial Lorentz factor ($\gamma_c = 5$) and magnetic pitch parameter fixed. These are categorized as resistive ResRMHD simulations. In the fourth simulation, \textit{Res0g10p1} (also listed in Table~\ref{table:simulation_run}), we increased the axial Lorentz factor to $\gamma_c = 10$. We have not included any explicit resistivity, thus treating it as an ideal RMHD run. Figure~\ref{fig:Res12_0g5_10p1_MFlines} shows the magnetic field line structures (colored in a blue-hot scale) overlaid with fluid density (in an orange-red scale) for all three simulations, captured at the respective times when kink instabilities are clearly developed. Panel (a) corresponds to the \textit{Res2g5p1} run at $t = 234 \, t_0$, panel (b) shows the \textit{Res1g5p1} run at $t = 246 \, t_0$, and panel (c) depicts the \textit{Res0g10p1} simulation at $t = 249 \, t_0$. Figure~\ref{fig:Res12_0g5_10p1_E_evolution} presents the time evolution of volume-averaged energy components within the jet, EM, magnetic, kinetic (toroidal and axial), and thermal—for these three simulations. All energy quantities are normalized to the code units. These plots help to characterize how the jet's internal dynamics and energy conversion processes vary with different resistivity levels and jet Lorentz factors, particularly in the presence of kink instabilities.

The third column of Figure~\ref{fig:Res12_0g5_10p1_E_evolution} illustrates the energy evolution for the simulation run \textit{Res0g10p1}. The overall behavior of the energy components in this simulation is qualitatively similar to that observed in the reference run \textit{Res0g5p1}, with the primary difference being the higher energy values. This similarity arises from both simulations being conducted under ideal RMHD conditions, differing only in the axial Lorentz factor. In the bottom panel of the third column, we observe a second phase of toroidal kinetic energy growth beginning around $t = 200 \, t_0$, which reaches its peak at $t = 258 \, t_0$, marked by the vertical black dash-dot line. During this phase, the thermal energy also increases and attains its peak at $t = 222 \, t_0$, indicated by the vertical maroon dash-dot line. This enhanced kinetic and thermal energy period coincides with a decline in magnetic and EM energies, suggesting active energy conversion via magnetic dissipation. Notably, this interval also corresponds to the development of kink instability, as depicted in Figure~\ref{fig:Res12_0g5_10p1_MFlines}(c), captured at $t = 249 \, t_0$.

Similarly, the first and second columns of Figure~\ref{fig:Res12_0g5_10p1_E_evolution} show the evolution of volume-averaged energy components for the simulations \textit{Res2g5p1} and \textit{Res1g5p1}, respectively. Both simulations exhibit a single prominent phase where the toroidal kinetic energy rises, reaches a peak, and subsequently declines. Thermal energy follows a similar trend during this phase, while magnetic energy shows a corresponding decrease, which is indicative of magnetic dissipation. For the \textit{Res2g5p1} run, the thermal energy peaks at $t = 234 \, t_0$, and the toroidal kinetic energy reaches its maximum at $t = 252 \, t_0$. The development of kink instability is observed between these two times, as illustrated in Figure~\ref{fig:Res12_0g5_10p1_MFlines}(a). Likewise, in the \textit{Res1g5p1} simulation, the thermal and toroidal kinetic energies peak at $t = 246 \, t_0$ and $t = 270 \, t_0$, respectively, with the onset of kink instability occurring in between, as shown in Figure~\ref{fig:Res12_0g5_10p1_MFlines}(b).

All four simulations show a phase evolution of the CDK instability, which is characterized by a decline in magnetic and EM energies followed by an increase in toroidal kinetic and thermal energy, indicating that an increase in turbulence may cause fast magnetic reconnection. The deformation of the original helical magnetic field lines indicates the onset of the kink mode. A comparative analysis among these simulations highlights the influence of resistivity and Lorentz factor on the instability onset and energetics. The temporal correlation between the peak of thermal and toroidal kinetic energy and magnetic structural distortion is consistent across every case, supporting the interpretation that turbulence-induced CDK facilitates reconnection in relativistic jets.

\subsection{Current sheet distribution} \label{subsec:CurrentSheetDist}
For the identification of current sheets in our simulated jet, we follow the methods described in the section \ref{Ident_currentsheet}. These current sheets form naturally as a result of the nonlinear growth of the CDK instability. As the kink instability grows, the magnetic field lines get distorted, resulting in strong current layers. These are the primary sites for reconnection, which leads to the conversion of magnetic energy into thermal and kinetic energy.

\begin{figure*}[ht!]
    \centering
    \includegraphics[width=1\linewidth]{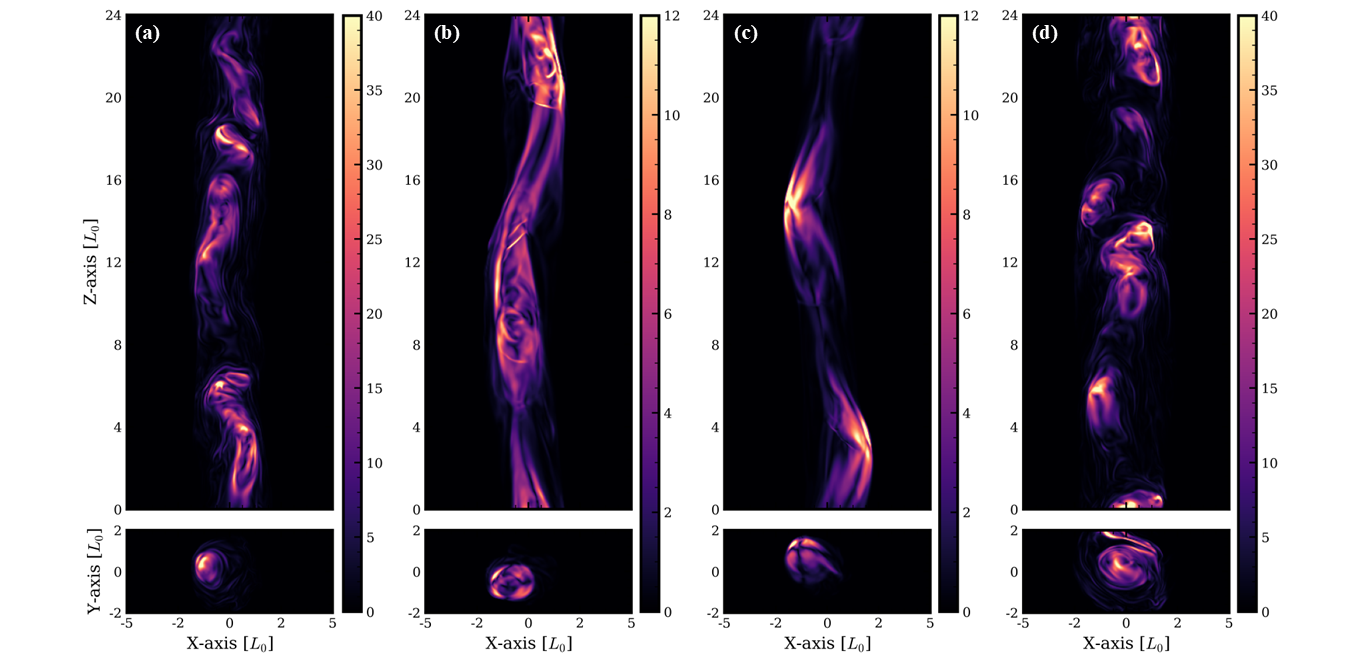}
    \caption{Current density profiles for four different simulation runs at specific time snapshots: \textbf{Res0g5p1} at $t = 213 \, t_0$, \textbf{Res2g5p1} at $t = 234 \, t_0$, \textbf{Res1g5p1} at $t = 246 \, t_0$, and \textbf{Res0g10p1} at $t = 258 \, t_0$. The color map used is \texttt{magma}, where brighter colors indicate higher current density. Each column shows two orthogonal slices of the current density magnitude $|\mathbf{J}|$ in the XZ- and XY-planes.}
    \label{fig:CurrentDensity}
\end{figure*}

Figure \ref{fig:CurrentDensity} shows eight subplots organized into four columns, corresponding to four different simulation runs: \textit{Res0g5p1}-(a), \textit{Res2g5p1}-(a), \textit{Res1g5p1}-(a), and \textit{Res0g10p1}-(a) (left to right). Each column presents two slices of the current density magnitude $|\mathbf{J}|$ in the XZ- and XY-planes at a time step corresponding to the nonlinear phase of the CDK instability. Here the time steps for the runs \textit{Res0g5p1}, \textit{Res2g5p1}, \textit{Res1g5p1}, and \textit{Res0g10p1} are at $t = 213 \, t_0$, $t = 234 \, t_0$, $t = 246 \, t_0$, and $t = 258 \, t_0$ respectively. The XZ-slices reveal elongated, sheet-like features along the jet axis, while the XY-slices highlight the azimuthal filamentation and twisting of current layers. These current sheets form as a result of the distortion of the magnetic field lines, driven by the kink instability. Their formation is temporally correlated with phases of enhanced magnetic energy dissipation and spatially associated with potential sites of particle acceleration. Thus, the current density profile is an important diagnostic tool for identifying locations of magnetic reconnection and understanding the underlying mechanics of energy conversion in relativistic jets. Notably, identifying and characterizing these current sheets, particularly their spatial extent and distribution, could have a significant impact on the observed variability timescales.

In magnetized plasmas, reconnection usually occurs in regions with strong current density, often forming thin, localized current sheets near magnetic null points. To identify these structures systematically, we used the {\tt Astrodendro} algorithm, which is designed to detect hierarchical structures in 3D data. As discussed earlier, this method requires two main input parameters: a threshold value for the current density (below which no structure is considered) and a minimum number of grid cells for a region to consider as an independent structure. For our analysis, we chose the average current density within the simulation volume as the threshold and set the minimum size to 27 grid cells, equivalent to a $3 \times 3 \times 3$ cube, assuming this is the smallest meaningful structure to detect. Using these settings, the algorithm identified local peaks in the current density profile, which we interpret as candidate current sheets that could be associated with magnetic reconnection.

To verify that the identified structures correspond to true magnetic reconnection sites, not just localized current concentrations arising from turbulence, we employed the ARD function from the {\tt LoRD} toolkit. This tool is specifically designed to detect magnetic reconnection regions by examining the topology of the magnetic field and quantifying reconnection metrics across the volume. When applied to our data, the ARD function independently identified reconnection sites. As with the dendrogram analysis, this method requires a threshold to filter out meaningful regions. In this case, we adopted the average electric field strength within the jet as the threshold criterion for reconnection identification.

Subsequently, we cross-referenced the current sheets identified via {\tt Astrodendro} ($S_{Ad}$) with the reconnection sites found by the {\tt LoRD} toolkit ($S_{L}$). The overlapping regions ($S_{Ad} \, \cap \, S_{L}$) between these two independent identification methods were considered reconnection-driven current sheets, which are the sites likely responsible for particle acceleration and localized energy conversion.

\begin{figure}[ht!]
    \centering
    \includegraphics[width=1\linewidth]{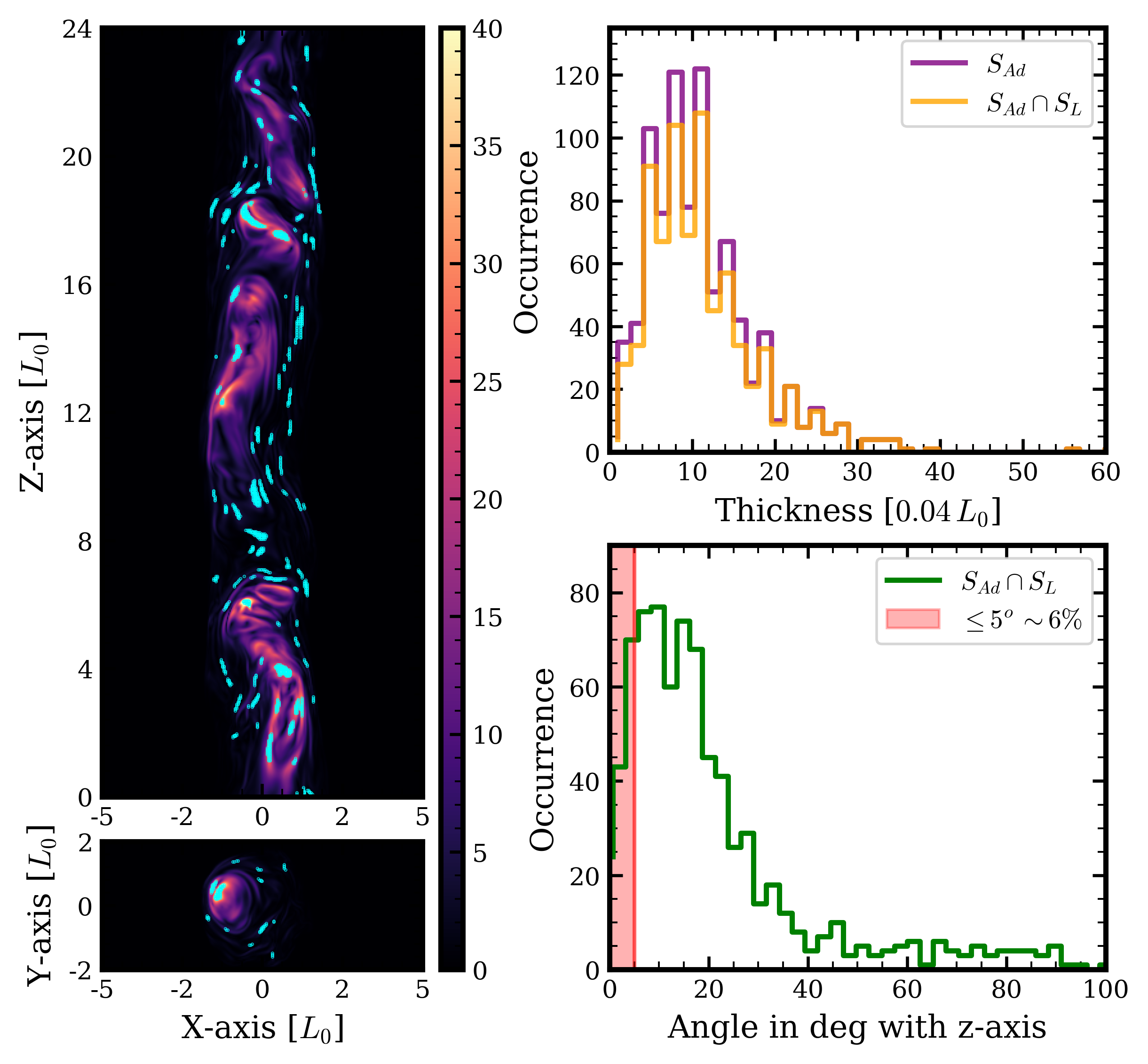}
    \caption{Column1: XZ- and XY- slice plot of current density profile overplotted with identified current sheets ($S_{Ad} \, \cap \, S_{L}$) from \textit{Astrodendro+LoRD} method (represented in cyan colour). Top right: Thickness distribution of identified current sheets form both \textit{Astrodendro} and \textit{Astrodendro+LoRD} method. Bottom left: Angular distribution of identified current sheets moving at different angle with Z-axis. This plot is for the simulation run \textbf{Res0g5p1} at $t = 213 \, \rm{L_0/c}$.}
    \label{fig:Res0g5p1_OD}
\end{figure}

Figure~\ref{fig:Res0g5p1_OD} provides a detailed analysis of the current sheets identified in the simulation run \textit{Res0g5p1} at $t = 213 \, t_0$, during the nonlinear phase of the CDK instability. The first column presents the spatial distribution of current sheets identified using the combined {\tt Astrodendro + LoRD} technique, overlaid on current density slices in the XZ and XY planes. The top-right panel shows the thickness distribution of current sheets ($S_{Ad}$), with results from the {\tt Astrodendro} method plotted in purple and those ($S_{Ad} \, \cap \, S_{L}$) from the combined approach in orange - both following a log-normal-like distribution (see Section~\ref{subsec:GeometricalDistribution} for a detailed discussion on the thickness and its calculation). Cross-matching analysis reveals that approximately $88\%$ of the $S_{Ad}$ initially identified by {\tt Astrodendro} coincide with reconnection regions detected by the {\tt LoRD toolkit}, supporting the robustness of our identification method. The bottom-right panel displays the angular distribution of the current sheets relative to the jet axis (z-axis).

\begin{figure*}[ht!]
    \centering
    \includegraphics[width=1\linewidth]{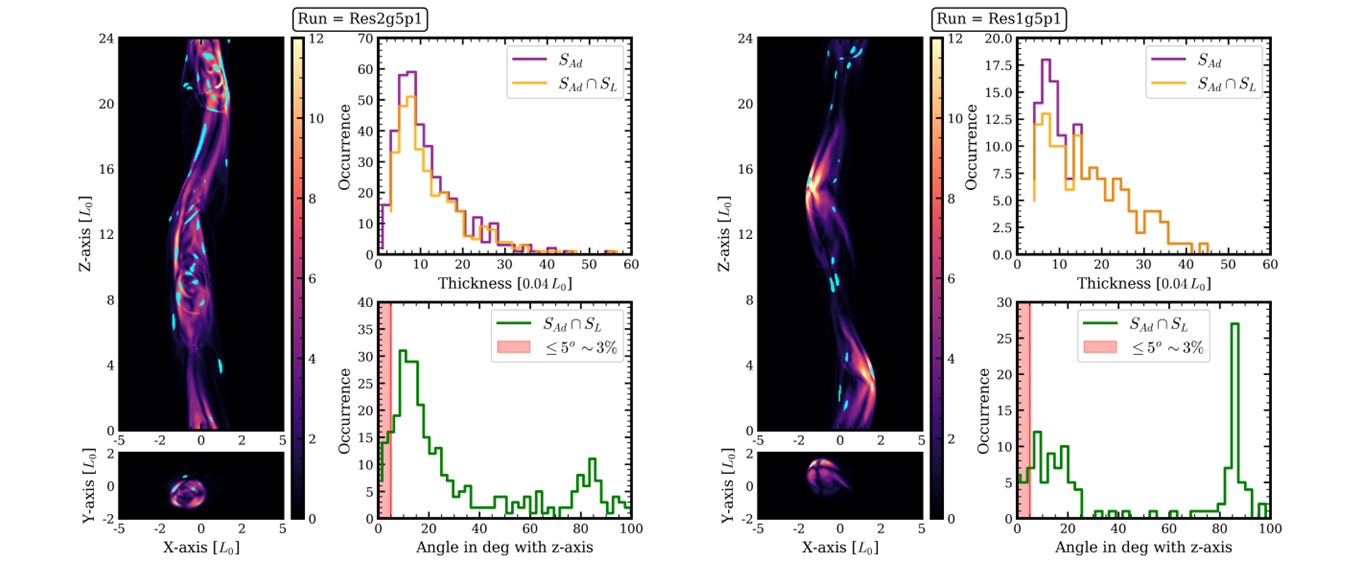}
    \caption{Spatial, thickness and angular distribution of identified current sheets for two different simulation runs; \textit{left:} \textbf{Res2g5p1} at $t = 234 \, t_0$ and \textit{right:} \textbf{Res1g5p1} at $t = 246 \, t_0$. For both the case the column 1 represents XZ- and XY- slice plot of current density profile overplotted with identified current sheets ($S_{Ad} \, \cap \, S_{L}$) from \textit{Astrodendro+LoRD} method (represented in cyan colour). Top right: Thickness distribution of identified current sheets form both \textit{Astrodendro} and \textit{Astrodendro+LoRD} method. Bottom left: Angular distribution of identified current sheets moving at different angle with Z-axis.}
    \label{fig:ResCombDistg5}
\end{figure*}

To understand how the properties of current sheets change with different resistivity and with a higher Lorentz factor, we applied our identification technique on three different simulation runs: \textit{Res2g5p1}, \textit{Res1g5p1}, and \textit{Res0g10p1}. Figure~\ref{fig:ResCombDistg5} and ~\ref{fig:NonResCombDistng10} show the current sheet structures identified using the {\tt Astrodendro+LoRD} algorithm, based on the 3D current density profiles. Figure~\ref{fig:ResCombDistg5} represents the distribution of identified current sheets for two different simulation runs; \textit{Res2g5p1} at $t = 234 \, t_0$ and \textit{Res1g5p1} at $t = 246 \, t_0$. Similarly, figure~\ref{fig:NonResCombDistng10} represents the distribution of identified current sheets for two different time steps for the simulation run; \textit{Res0g10p1} at $t = 240 \, t_0$ and at $t = 258 \, t_0$. The first column in these four plots represents one of the simulation runs, with the top and bottom panels showing XZ and XY slices, respectively. The cyan-colored volumetric regions correspond to the identified current sheets that are responsible for magnetic reconnection. For \textit{Res2g5p1} simulation run which is with explicit resistivity $10^{-4}$, we observe a reduced number of identified current sheets compared to the reference run. Furthermore, the number of current sheets decreases as the level of resistivity increases. This trend is expected, as the presence of explicit resistivity acts to smooth out magnetic field gradients, thereby suppressing instabilities. Consequently, the jet becomes less turbulent, leading to fewer localized regions of enhanced current density. In contrast, the third column represents the simulation with a higher Lorentz factor and no explicit resistivity (i.e., only numerical resistivity is present). In this case, a significantly larger number of current sheets are observed. However, these structures are generally smaller in size compared to those in the resistive simulations. This distinct difference in morphology and distribution of current sheets will be examined in detail in the following subsection.

\subsection{Distribution of geometrical properties of current sheets}  \label{subsec:GeometricalDistribution}
As discussed previously, the size and orientation of plasmoids, structures formed within current sheets as a result of magnetic reconnection, relative to the observer, play a crucial role in shaping the observed flux distribution and variability timescales. To isolate these structures, we applied a combined approach using the {\tt Astrodendro} algorithm and the ARD module from the {\tt LoRD} toolkit. This integrated method allows us to systematically identify volumetric current sheet regions that spatially overlap with reconnection zones, thereby flagging them as likely sites of plasmoid formation and particle acceleration.

A key advantage of using the dendrogram-based {\tt Astrodendro} technique is that it enables a hierarchical classification of individual current sheets, allowing us to independently study their spatial distribution and morphological properties. Given that our study focuses on blazar-like sources, where the observed variability is primarily governed by structures aligned along the jet axis (z-axis), we define the thickness of a current sheet as its extent along the z-direction. To produce the thickness distribution, we compute the z-elongation for each identified current sheet in both methods, {\tt Astrodendro} and the combined {\tt Astrodendro+LoRD}.

The angle at which the current sheets propagate relative to the jet axis is another significant element affecting the observed variability. This angle plays a role in determining the effective Doppler boosting of the emitted radiation. Since we focus on blazar-type sources, where the line of sight is closely aligned with the jet axis (z-axis), even small variations in the propagation angle can significantly impact the observed flux and variability timescale. To quantify this effect, we first compute the bulk velocity vector of each identified current sheet using the local plasma flow velocities averaged over its volume. From this velocity vector, we determine the angle it makes with the z-axis. This angle serves as a proxy for the orientation of the current sheet motion with respect to the observer's line of sight. The resulting angle distribution provides key insights into the kinematic properties of the reconnection-driven structures and their role in shaping the observed variability characteristics of relativistic jets.
\begin{figure*}[ht!]
    \centering
    \includegraphics[width=1\linewidth]{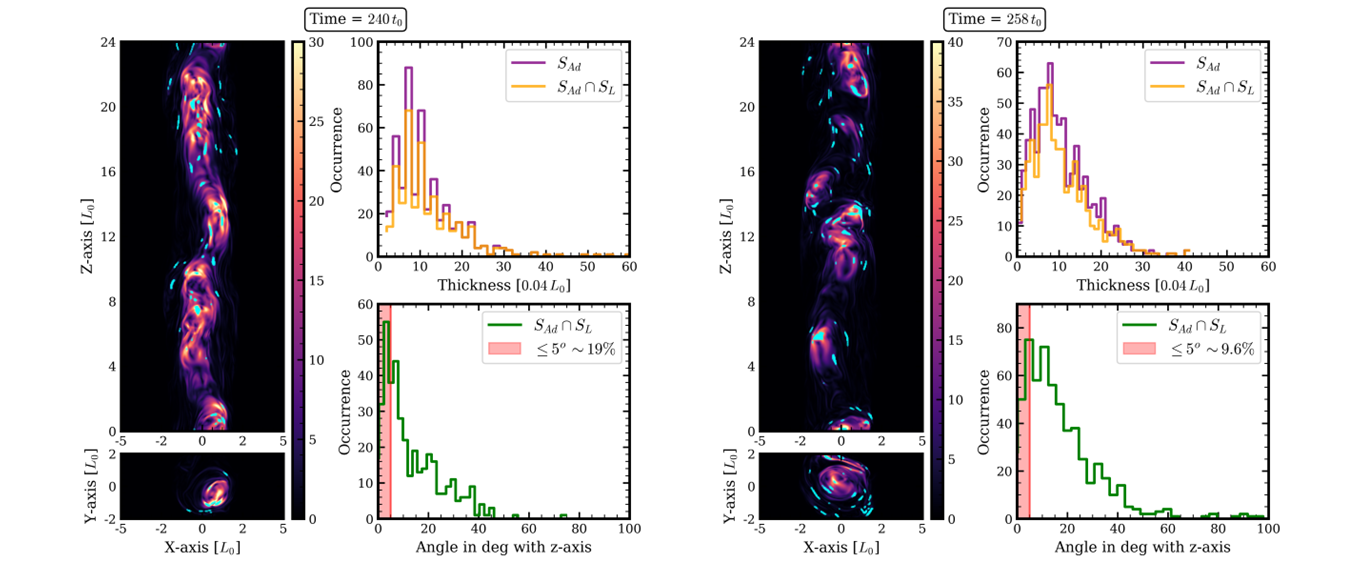}
    \caption{Spatial, thickness and angular distribution of identified current sheets for two different time steps for the simulation run; \textbf{Res0g10p1} \textit{left:} is at $t = 240 \, t_0$ and \textit{right:} is at $t = 258 \, t_0$. For both the case the column 1 represents XZ- and XY- slice plot of current density profile overplotted with identified current sheets ($S_{Ad} \, \cap \, S_{L}$) from \textit{Astrodendro+LoRD} method (represented in cyan colour). Top right: Thickness distribution of identified current sheets form both \textit{Astrodendro} and \textit{Astrodendro+LoRD} method. Bottom left: Angular distribution of identified current sheets moving at different angle with Z-axis.}
    \label{fig:NonResCombDistng10}
\end{figure*}

Figure~\ref{fig:Res0g5p1_OD} illustrates the thickness (top-right subplot) and angular (bottom-right subplot) distributions of the identified current sheets for the reference simulation run \textit{Res0g5p1} at $t = 213, t_0$. The thickness distribution exhibits a log-normal shape, consistent with the observed flux distributions in blazars when the emission originates from a multiplicative process, such as magnetic reconnection \citep{Abramowski2010}. This connection will be further explored in the next section. In the angular distribution plot, we find that only a small fraction of current sheets (approximately $6.4 \%$) are oriented within $5^{\circ}$ of the jet axis (z-axis), which closely aligns with the observer's line of sight in blazar-type sources. Beyond this angle, the distribution does not exhibit a clear pattern, suggesting a more random orientation of the remaining current sheets. As previously discussed, plasmoids formed within these sheets that move randomly with respect to the observer's line of sight contribute to a slowly varying envelope in the observed light curve. In contrast, those aligned within a narrow cone around the jet axis can produce sharp, transient flares superimposed on this envelope. Hence, the current sheets within $5^{\circ}$ are expected to be the primary contributors to rapid flux variability, while those at larger angles contribute to the underlying gradual emission.

\begin{deluxetable*}{ c  c  c  c  c  c }
\renewcommand{\arraystretch}{1.2}
\tablecaption{Summary of current sheet properties for different simulation runs. The table lists the total number of identified current sheets using the {\tt Astrodendro+LoRD} methods, the percentage of sheets oriented within $5^{\circ}$ of the $z$-axis (i.e., close to the observer's line of sight, contributing to fast flares via Doppler boosting), the maximum Alfv\'en velocity and the maximum Doppler factor among these aligned current sheets.\label{table:result_final}}
\tablehead{
\colhead{\textbf{Run Name}} & 
\colhead{\textbf{Time step}} & 
\colhead{\textbf{No. of identified CS}} & 
\colhead{\textbf{$\%$ of CS within $5^{\circ}$}} &
\colhead{\textbf{$V_{alf, max}$ within $5^{\circ}$}} &
\colhead{\textbf{$\delta_{max}$ within $5^{\circ}$}}
}
\startdata
Res0g5p1 & $198 \, t_0$ & $537$ & $7.3$ & $0.92 \, c$ & $7.92$ \\ \hline
Res0g5p1 & $213 \, t_0$ & $777$ & $6.4$ & $0.83 \, c$ & $3.50$ \\ \hline
Res2g5p1 & $234 \, t_0$ & $327$ & $3.0$ & $0.86 \, c$ & $3.57$ \\ \hline
Res1g5p1 & $246 \, t_0$ & $154$ & $5.3$ & $0.83 \, c$ & $3.82$ \\ \hline
Res0g10p1 & $240 \, t_0$ & $399$ & $19.0$ & $0.98 \, c$ & $9.51$ \\ \hline
Res0g10p1 & $258 \, t_0$ & $596$ & $9.6$ & $0.93 \, c$ & $5.30$ \\
\enddata
\end{deluxetable*}

\begin{figure*}[ht!]
    \centering
    \includegraphics[width=1\linewidth]{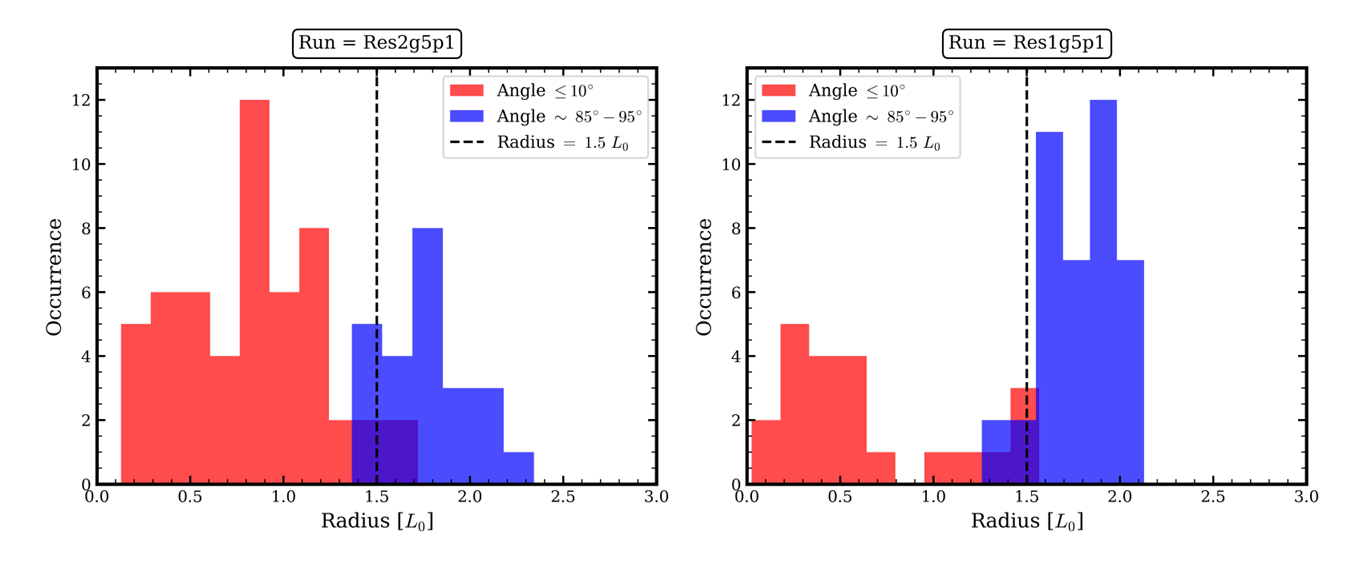}
    \caption{Radial distribution of current sheets in \textit{Res2g5p1} and \textit{Res1g5p1} for two orientation ranges: within $10^\circ$ of the jet axis (shown in red) and within $85^\circ$ – $95^\circ$ (shown in blue). The vertical dashed line at radius $r = 1.5 \, L_0$ marks the jet boundary.}
    \label{fig:Res_RadDist}
\end{figure*}

Similarly, Figure~\ref{fig:ResCombDistg5} and ~\ref{fig:NonResCombDistng10} present the thickness and angular distributions of current sheets for three other simulation runs: \textit{Res2g5p1} at $t = 234 \, t_0$, \textit{Res1g5p1} at $t = 246 \, t_0$, and \textit{Res0g10p1} at $t = 240 \, t_0$ and $t = 258 \, t_0$. Similar to the reference run, the thickness distributions in all cases exhibit a log-normal-like behavior. However, a noticeable trend is the decreasing number of current sheets with increasing resistivity. As reported in Table~\ref{table:result_final}, for the simulations with axial Lorentz factor $\gamma_c = 5$, the number of current sheets identified are $777$ for \textit{Res0g5p1} at $t = 213 , t_0$, $327$ for \textit{Res2g5p1}, and $154$ for \textit{Res1g5p1}, clearly showing a reduction with increasing explicit resistivity. For \textit{Res0g10p1}, we perform our analysis at two different time steps: the onset of the instability phase at $t = 240 \, t_0$ and a later stage at $t = 258 \, t_0$, and a similar time-dependent increase in the number of current sheets is also observed in the \textit{Res0g5p1} case, where the count rises from $537$ at $t = 198 \, t_0$ to $777$ at $t = 213 \, t_0$. As shown in Table~\ref{table:result_final}, the instability is higher at later times, which leads to the formation of a larger number of current sheets.

Figure~\ref{fig:ResCombDistg5} and ~\ref{fig:NonResCombDistng10} also illustrate the angular distribution of these current sheets. The red shaded region marks the angle that is made by current sheets less than $5^{\circ}$ around the $z$-axis. As discussed earlier, plasmoids formed in sheets within this narrow region can lead to rapid flares in observed blazar light curves due to enhanced Doppler boosting along the line of sight. Conversely, the slowly varying envelope in the light curve arises from plasmoids in current sheets that deviate more than $5^{\circ}$ from the $z$-axis. The proportion of current sheets falling within the $5^{\circ}$ cone, as listed in Table~\ref{table:result_final}, are $3.0\%$ and $5.3\%$ for \textit{Res2g5p1} and \textit{Res1g5p1}, respectively. For the simulations \textit{Res0g5p1} and \textit{Res0g10p1}, we performed the analysis at two different time steps during the kink instability phase. In the case of \textit{Res0g5p1}, the percentage of current sheets within $5^{\circ}$ of the jet axis decreases from $7.3\%$ at $t = 198 \, t_0$ to $6.4\%$ at $t = 213 \, t_0$. Similarly, for \textit{Res0g10p1}, it decreases from $19.0\%$ at $t = 240 \, t_0$ to $9.6\%$ at $t = 258 \, t_0$. After comparison at two different time steps, we observe a decrease in the number of sheets oriented within $5^{\circ}$ of the jet axis at a later time, indicating that such favorable alignments become less common as the kink instability evolves. This trend suggests that strongly Doppler-boosted events, which are responsible for producing fast flares, may be less prevalent in the later stages of kink-driven instability. The change in both orientation and thickness distributions over time highlights the dynamic nature of current sheet evolution and the transient window during which reconnection-driven flaring may be observationally detectable.

Furthermore, we calculate the Alfvén velocity, $V_{alf}$, which is the characteristic speed at which plasmoids, formed within the current sheets (CSs) located within $5^{\circ}$ of the jet axis, propagate in the jet frame. For each CS in this angular range, we determine the maximum Alfvén velocity, $V_{alf, max}$, attainable in our simulations. From $V_{alf,max}$, we then estimate the corresponding maximum Doppler factor, $\delta_{max}$, using the standard relativistic beaming relation:

\begin{equation} \label{eq:Doppler}
    \delta = \frac{1}{\Gamma (1 - \beta cos\theta)}
\end{equation}
where $\Gamma$ is the bulk Lorentz factor of the jet, $\beta$ is the Alfvén velocity in units of $c$, and $\theta$ is the inclination angle to the observer (taken here as $5^{\circ}$). The resulting $\delta_{max}$ values are listed in Table~\ref{table:result_final}, providing an estimate of the highest apparent boosting that reconnection-driven plasmoids could achieve in our simulated jets.

Comparing the angular distributions across all four simulations, the runs with explicit resistivity (\textit{Res2g5p1} and \textit{Res1g5p1}) show an additional prominent peak near $\sim 85^{\circ}$. In contrast, simulations without explicit resistivity (\textit{Res0g5p1} and \textit{Res0g10p1}) do not display such a feature. To investigate this, we plot the radial distribution of current sheets in Figure~\ref{fig:Res_RadDist} for two types of orientations: sheets with angles within $10^{\circ}$ of the jet axis (red) and those within $85^{\circ}$–$95^{\circ}$ (blue), for two resistive run cases (\textit{Res2g5p1} and \textit{Res1g5p1}). We observe that current sheets oriented within $10^{\circ}$ are primarily confined within the jet ($r < 1.5 \, L_0$), while those oriented within $85^{\circ}-95^{\circ}$ are mainly located outside the jet. This suggests that the additional peak in angular distribution in the resistive runs originates from the outer jet boundary, likely due to turbulence driven by interactions with the ambient medium. These sheets move nearly perpendicular to the line of sight and are unlikely to contribute significantly to the observed blazar light curves.

\section{Discussion and Summary} \label{sec:Discussion}
In this study, we have performed a series of high-resolution 3D RMHD simulations of AGN jets, covering both ideal and resistive cases, to investigate the formation and evolution of current sheets and their role in driving fast flares via magnetic reconnection. Our simulations encompass a variety of physical conditions, varying both resistivity and axial Lorentz factor, enabling a systematic investigation of how these parameters affect jet dynamics and the morphology of reconnection-driven current sheets.

We used the {\tt \string PLUTO} code to simulate the evolution of relativistic jets up to the onset of turbulence. Across all simulation runs, magnetic energy was gradually converted into kinetic energy as turbulence developed. This energy redistribution suggests that magnetic energy is efficiently converted into plasma motion and heating, consistent with theoretical expectations of the kink instability leading to reconnection-driven energy dissipation \citep{Striani2016, Zhdankin2017, Mattia2023, Mattia2024}. Visualizations revealed kink instabilities as the primary driver of this transition, which may facilitate magnetic reconnection \citep{Mizuno2012, Bromberg2019}. We analyzed the 3D current density profile to identify reconnection sites and applied the {\tt Astrodendro} method to extract current sheets from local maxima. We then validated these structures using the ARD function from the {\tt LoRD} toolkit, confirming that a subset of the identified current sheets are actively undergoing magnetic reconnection.

Theoretical framework such as the jets-in-jet model \citep{Giannios2009} provides a compelling explanation for the observed short-timescale variability. In this model, magnetic reconnection in the jet generates small-scale plasmoids (mini-jets) that can move relativistically within the broader jet flow. When these plasmoids are aligned with the observer's line of sight, the resulting emission is strongly Doppler-boosted, leading to intense and rapid flares in observed $\gamma$-ray light curves. Our analysis of thickness and angular distributions of identified current sheets supports this model: a non-negligible fraction of reconnection sites are oriented within $5^\circ$ of the jet axis and their length scale along our line of sight, making them likely candidates for producing such flares. However, it is essential to emphasize the limitations of our current simulations. Due to resolution constraints, we cannot resolve individual plasmoids within the current sheets. Our interpretation thus assumes that these thin, jet-aligned current sheets would fragment into plasmoids moving at relativistic speeds, comparable to the local Alfv\'en velocity. While we do not directly calculate the Doppler factor of plasmoids in this work, the inferred orientation and dynamics of these sheets suggest that they could produce flares with Doppler-boosting exceeding the typical bulk Lorentz factor of the jet, potentially explaining fast variability in blazar light curves. This picture is broadly consistent with particle-in-cell (PIC) simulations, which confirm that kink instabilities in relativistic jets can drive efficient nonthermal particle acceleration \citep{Alves2018}.

In simulations with explicit resistivity (e.g., \textit{Res2g5p1}, \textit{Res1g5p1}), we observe a lesser number of current sheets compared to the ideal RMHD case (\textit{Res0g5p1}), indicating that reconnection becomes more localized as resistivity increases. This trend is quantitatively evident in table~\ref{table:result_final}, which presents the total number of current sheets and the percentage of those aligned within $5^\circ$ of the jet axis. For all simulations with an axial Lorentz factor of $\gamma_c = 5$, we find that only a small fraction of current sheets are aligned within $5^\circ$, suggesting that such geometrically favorable configurations are relatively rare. This rarity is significant because current sheets aligned along the line of sight are expected to produce strongly Doppler-boosted emission, manifesting as fast flares superimposed on a slowly varying envelope. The low occurrence rate of such aligned sheets implies that these fast flares are likely rare and discrete, aligning with the observational profiles of minute-scale variability in blazars \citep{Albert2007, Aharonian2007, Shukla2020}. Interestingly, we observe a slightly higher percentage of aligned current sheets in the case of \textit{Res0g10p1} (with $\gamma_c = 10$). However, when comparing two different snapshots during the kink instability phase at $t = 240 \, t_0$ and $t = 258 \, t_0$, the number of aligned sheets decreases at the later time. This trend suggests that even in high Lorentz factor jets, such reconnection events with favorable alignment become rare as the kink instability evolves. This time-dependent reduction may reflect the dissipation and reorganization of magnetic structures (e.g., current sheets and plasmoids) as the jet transitions toward a more turbulent state.

In addition to the dominant axis-aligned current sheets, our resistive simulations reveal an excess of nearly transverse sheets ($\sim85^{\circ}$) primarily located at the jet boundary. While these sheets are unlikely to contribute directly to fast $\gamma$-ray flares, since they move nearly perpendicular to the line of sight, their presence highlights the role of jet–environment interactions in shaping the current sheets population. Such boundary-driven turbulence may leave observable imprints on the large-scale jet, as suggested by the peculiar radio morphology of the TeV source IC 310 \citep{Kadler2012} due to jet–ambient medium interaction. This suggests that transverse current sheets, although not dominant for high-energy variability, may still have observable imprints on the radio morphology of AGN jets.

The thickness distribution of the identified current sheets, shown in figures~\ref{fig:ResCombDistg5}, and \ref{fig:NonResCombDistng10}, consistently exhibits a log-normal-like profile. This characteristic shape suggests a hierarchical formation of current sheets, likely driven by turbulence within the jet. Such distributions have been associated with stochastic reconnection environments, where turbulence fragments larger sheets into smaller structures over time \citep{Lazarian1999, Zhdankin2013, GouveiaDalPino2014, Kowal2017}. Moreover, this log-normal behavior is also found in the $\gamma$-ray flux distribution observed in blazars \citep{Abramowski2010}. This supports the idea that the emission arises from a multiplicative process, such as that driven by magnetic reconnection and plasmoid formation dynamics. The consistency of this pattern in our simulations supports the idea that magnetic reconnection, driven by turbulence, could play a key role in powering the fast and powerful flares observed in blazars.

While our current analysis focuses on the geometric and statistical properties of current sheets, a promising future direction is to perform power spectral density (PSD) analysis of reconnection activity or synthetic light curves. This could reveal whether variability patterns in the simulations resemble the flickering power-law noise observed in blazars, offering more profound insight into the link between current sheet dynamics and high-energy emission. Although our simulations capture essential features of the jet-in-jet model, they still have certain limitations. We adopt a simplified resistivity model, and radiation feedback is not yet included. Radiative losses will be included in future research to improve the accuracy of flare energetics and spectrum predictions. Linking these results to artificial light curves and polarization signatures would provide useful comparisons with current and upcoming high-energy observations. Nevertheless, we emphasize that the internal structure of reconnection regions is not resolved within our MHD framework. Future work should incorporate kinetic effects, as in \citet{Nalewajko2018}, and explore the role of the Hall term, which has been shown to strongly influence reconnection dynamics in PIC studies \citep{Alves2018}.

In summary, our high-resolution 3D RMHD simulations with both ideal and resistive cases reveal the development of the CDK instability within relativistic AGN jets, leading to magnetic field deformation and turbulence that facilitate the fast magnetic reconnection. We introduce a novel current sheet identification technique that enables robust characterization of their geometry and statistical properties. The resulting current sheets exhibit a broad distribution of orientations and lengths, with a subset forming along the jet axis, consistent with the jet-in-jet scenario to explain fast blazar variability. The inferred dynamics and alignment of these structures suggest that they can produce relativistically moving plasmoids with enhanced Doppler boosting, offering a viable mechanism for the rapid flare on top of the slowly varying envelope in blazar light curves at very high energy.

\section*{Acknowledgment}
AS acknowledges the Department of Science and Technology (DST), India, for financial support under grant number CRG/2022/009332. KM and BV acknowledge support from the Deutsche Forschungsgemeinschaft (DFG, German Research Foundation) as part of the DFG Research Unit FOR5195 – project number 443220636. BV acknowledges the support from the Max Planck Partner group grant that was established at IIT Indore. This research made use of the high-performance computing facilities at IIT Indore and the Max Planck Group’s supercomputing cluster Raven. This research made use of astrodendro, a Python package to compute dendrograms of astronomical data \url{http://www.dendrograms.org/}. Also, this work made use of Astropy: \url{http://www.astropy.org}, a community-developed core Python package and an ecosystem of tools and resources for astronomy.

\appendix
\section{Initial setup} \label{Appendix:InitialSetup}
For the initial setup of the plasma column, our investigation begins with a consideration of a 3D initial configuration of a force-free field \citep{Bodo2019} in the cylindrical coordinate system. There is no radial component of velocity and magnetic field. It only contains poloidal ($v_z, \, B_z$) and toroidal ($v_{\phi}, \, B_{\phi}$) components. The field strength in cylindrical coordinates is defined as:

\begin{equation} \label{eqA1}
    \begin{split}
    &H = \frac{H_c}{r}\sqrt{\left[1-exp\left(-\frac{r^4}{a^4}\right)\right]}\\
    \end{split}
\end{equation}
Here, $H^2=B_{\phi}^2 - E_r^2$, with $E_r = v_zB_{\phi}-v_{\phi}B_z$. The quantity $H_c$ denotes the value of $H$ along the jet axis, $a = 0.6r_j$ defines the magnetization radius, and the jet radius is set to $r_j = 1$. The magnetic field configuration is characterized by the pitch parameter $P = \frac{rB_z}{B_{\phi}}$. Since the magnetic structure of astrophysical jets is not directly observable, we adopt an idealized radial profile where the vertical current density peaks along the axis and is confined within the magnetization radius $a$. The Lorentz factor, determined solely by the axial velocity component, is given by:

\begin{equation} \label{eqA2}
    \gamma_z(r) = 1 + \frac{\gamma_c - 1}{cosh(r/r_j)^6},
\end{equation}
and the azimuthal component of velocity is taken as,
\begin{equation} \label{eqA3}
    \gamma^2v_\phi^2 = (r\Omega_c\gamma_c)^2exp\left(-\frac{r^4}{a^4}\right),
\end{equation}
where $\gamma$ is the total Lorentz factor, $\Omega_c$ is the angular velocity, and $\gamma_c = (1 - v_c^2)^{-1/2}$ is the Lorentz factor associated with the axial velocity on the central axis, with $v_c = v_z(0)$. From Equation~(\ref{eqA3}), the expression for $v_{\phi}$ is obtained as:

\begin{equation} \label{eqA4}
    v_\phi^2 = \left(\frac{r\Omega_c\gamma_c}{\gamma_z}\right)^2\left[1+r^2\Omega_c^2\gamma_c^2 exp\left(-\frac{r^4}{a^4}\right)\right]^{-1}exp\left(-\frac{r^4}{a^4}\right).
\end{equation}

The only remaining non-trivial equation is the radial component of the momentum equation, which simplifies in the zero-pressure limit to:

\begin{equation} \label{eqA5}
    \rho \gamma^2 v_\phi^2 = \frac{1}{2r}\frac{d(r^2H^2)}{dr} + \frac{r}{2}\frac{dB_z^2}{dr}.
\end{equation}

From equation (\ref{eqA1}), (\ref{eqA3}), and (\ref{eqA5}), we get the $B_z$ profile as:

\begin{equation} \label{eqA6}
    B_z^2 = B_{zc}^2 - (1-\alpha)\frac{H_c^2\sqrt{\pi}}{a^2}erf\left(\frac{r^2}{a^2}\right),
\end{equation}
where $\mathrm{erf}$ denotes the error function, and the parameter $\alpha = \frac{\rho \gamma_c^2 \Omega_c^2 a^4}{2H_c^2}$ quantifies the strength of rotation ($\alpha = 0$ for no rotation, and $\alpha = 1$ for pure rotation). Substituting the expressions for $B_z$, $v_z$, and $v_{\phi}$ into the definition of $H$, we obtain a quadratic equation for $B_{\phi}$, whose solution is:

\begin{equation} \label{eqA7}
    B_{\phi} = \frac{-v_{\phi} v_z B_z \mp \sqrt{v_{\phi}^2 B_z^2 + H^2(1-v_z^2)}}{1-v_z^2}.
\end{equation}

The negative branch is chosen to ensure that $B_{\phi}$ and $v_{\phi}$ have opposite signs, consistent with the acceleration mechanism proposed by \citet{Blandford1982}. Our configuration is specified by the absolute value of the pitch on the axis, $P_c$, the axial Lorentz factor $\gamma_c$, and the ratio of matter to magnetic energy density, $M_a^2$.

\begin{equation} \label{eqA8}
    P_c \equiv \left|\frac{rB_z}{B_{\phi}}\right|_{r=0}, \qquad M_a^2 \equiv \frac{\rho\gamma_c^2}{\left<\textbf{B}^2\right>}
\end{equation}
where, $\left<\textbf{B}^2\right> = \int_{0}^{r_j} (B_z^2 + B_{\phi}^2), r ,dr\big/\int_{0}^{r_j} r ,dr$ is the radially averaged magnetic energy density. The parameter $M_a$ is related to the standard magnetization parameter.

\bibliography{bib}{}

@Article{Kniffen1993,
  author    = {Kniffen, D. A. and Bertsch, D. L. and Fichtel, C. E. and Hartman, R. C. and Hunter, S. D. and Kanbach, G. and Kwok, P. W. and Lin, Y. C. and Mattox, J. R. and Mayer-Hasselwander, H. A. and Michelson, P. F. and von Montigny, C. and Nolan, P. L. and Pinkau, K. and Schneid, E. and Sreekumar, P. and Thompson, D. J.},
  journal   = {The Astrophysical Journal},
  title     = {Time variability in the gamma-ray emission of 3C 279},
  year      = {1993},
  issn      = {1538-4357},
  month     = jul,
  pages     = {133},
  volume    = {411},
  doi       = {10.1086/172813},
  publisher = {American Astronomical Society},
}

@Article{Shukla2018,
  author    = {Shukla, A. and Mannheim, K. and Patel, S. R. and Roy, J. and Chitnis, V. R. and Dorner, D. and Rao, A. R. and Anupama, G. C. and Wendel, C.},
  journal   = {The Astrophysical Journal Letters},
  title     = {Short-timescale γ-Ray Variability in CTA 102},
  year      = {2018},
  issn      = {2041-8213},
  month     = feb,
  number    = {2},
  pages     = {L26},
  volume    = {854},
  doi       = {10.3847/2041-8213/aaacca},
  publisher = {American Astronomical Society},
}

@Article{Aharonian2007,
  author    = {Aharonian, F. and Akhperjanian, A. G. and Bazer-Bachi, A. R. and Behera, B. and Beilicke, M. and Benbow, W. and Berge, D. and Bernlöhr, K. and Boisson, C. and Bolz, O. and Borrel, V. and Boutelier, T. and Braun, I. and Brion, E. and Brown, A. M. and Bühler, R. and Büsching, I. and Bulik, T. and Carrigan, S. and Chadwick, P. M. and Clapson, A. C. and Chounet, L.-M. and Coignet, G. and Cornils, R. and Costamante, L. and Degrange, B. and Dickinson, H. J. and Djannati-Ataï, A. and Domainko, W. and Drury, L. O’C. and Dubus, G. and Dyks, J. and Egberts, K. and Emmanoulopoulos, D. and Espigat, P. and Farnier, C. and Feinstein, F. and Fiasson, A. and Förster, A. and Fontaine, G. and Funk, Seb. and Funk, S. and Füssling, M. and Gallant, Y. A. and Giebels, B. and Glicenstein, J. F. and Glück, B. and Goret, P. and Hadjichristidis, C. and Hauser, D. and Hauser, M. and Heinzelmann, G. and Henri, G. and Hermann, G. and Hinton, J. A. and Hoffmann, A. and Hofmann, W. and Holleran, M. and Hoppe, S. and Horns, D. and Jacholkowska, A. and de Jager, O. C. and Kendziorra, E. and Kerschhaggl, M. and Khélifi, B. and Komin, Nu. and Kosack, K. and Lamanna, G. and Latham, I. J. and Le Gallou, R. and Lemière, A. and Lemoine-Goumard, M. and Lenain, J.-P. and Lohse, T. and Martin, J. M. and Martineau-Huynh, O. and Marcowith, A. and Masterson, C. and Maurin, G. and McComb, T. J. L. and Moderski, R. and Moulin, E. and de Naurois, M. and Nedbal, D. and Nolan, S. J. and Olive, J-P. and Orford, K. J. and Osborne, J. L. and Ostrowski, M. and Panter, M. and Pedaletti, G. and Pelletier, G. and Petrucci, P.-O. and Pita, S. and Pühlhofer, G. and Punch, M. and Ranchon, S. and Raubenheimer, B. C. and Raue, M. and Rayner, S. M. and Renaud, M. and Ripken, J. and Rob, L. and Rolland, L. and Rosier-Lees, S. and Rowell, G. and Rudak, B. and Ruppel, J. and Sahakian, V. and Santangelo, A. and Saugé, L. and Schlenker, S. and Schlickeiser, R. and Schröder, R. and Schwanke, U. and Schwarzburg, S. and Schwemmer, S. and Shalchi, A. and Sol, H. and Spangler, D. and Stawarz, Ł. and Steenkamp, R. and Stegmann, C. and Superina, G. and Tam, P. H. and Tavernet, J.-P. and Terrier, R. and van Eldik, C. and Vasileiadis, G. and Venter, C. and Vialle, J. P. and Vincent, P. and Vivier, M. and Völk, H. J. and Volpe, F. and Wagner, S. J. and Ward, M. and Zdziarski, A. A.},
  journal   = {The Astrophysical Journal},
  title     = {An Exceptional Very High Energy Gamma-Ray Flare of PKS 2155-304},
  year      = {2007},
  issn      = {1538-4357},
  month     = jul,
  number    = {2},
  pages     = {L71--L74},
  volume    = {664},
  doi       = {10.1086/520635},
  publisher = {American Astronomical Society},
}

@Article{Aartsen2018,
  author    = {Aartsen, Mark and Ackermann, Markus and Adams, Jenni and Aguilar, Juan Antonio and Ahlers, Markus and Ahrens, Maryon and Al Samarai, Imen and Altmann, David and Andeen, Karen and Anderson, Tyler and Ansseau, Isabelle and Anton, Gisela and Argüelles, Carlos and Arsioli, Bruno and Auffenberg, Jan and Axani, Spencer and Bagherpour, Hadis and Bai, Xinhua and Barron, Jared and Barwick, Steve and Baum, Volker and Bay, Ryan and Beatty, James and Becker, Karl Heinz and Becker Tjus, Julia and BenZvi, Segev and Berley, David and Bernardini, Elisa and Besson, David and Binder, Gary and Bindig, Daniel and Blaufuss, Erik and Blot, Summer and Bohm, Christian and Boerner, Mathis and Bos, Fabian and Boeser, Sebastian and Botner, Olga and Bourbeau, Etienne and Bourbeau, James and Bradascio, Federica and Braun, Jim and Brenzke, Martin and Bretz, Hans-Peter and Bron, Stephanie and Brostean-Kaiser, Jannes and Burgman, Alexander and Busse, Raffaela and Carver, Tessa and Cheung, Edward and Chirkin, Dmitry and Christov, Asen and Clark, Ken and Classen, Lew and Coenders, Stefan and Collin, Gabriel and Conrad, Janet and Coppin, Paul and Correa, Pablo and Cowen, Doug and Cross, Robert and Dave, Pranav and Day, Melanie and de André, Joao Pedro A. M. and De Clercq, Catherine and Delaunay, James and Dembinski, Hans and DeRidder, Sam and Desiati, Paolo and de Vries, Krijn and DeWasseige, Gwenhael and DeWith, Meike and DeYoung, Ty and Díaz-Vélez, Juan Carlos and Di Lorenzo, Vincenzo and Dujmovic, Hrvoje and Dumm, Jonathan and Dunkman, Matt and Dvorak, Emily and Eberhardt, Benjamin and Ehrhardt, Thomas and Eichmann, Bjorn and Eller, Philipp and Evenson, Paul and Fahey, Sam and Fazely, Ali and Felde, John and Filimonov, Kirill and Finley, Chad and Flis, Samuel and Franckowiak, Anna and Friedman, Elizabeth and Fritz, Alexander and Gaisser, Tom and Gallagher, Jay and Gerhardt, Lisa and Ghorbani, Kevin and Giommi, Paolo and Glauch, Theo and Gluesenkamp, Thorsten and Goldschmidt, Azriel and Gonzalez, Javier and Grant, Darren and Griffith, Zachary and Haack, Christian and Hallgren, Allan and Halzen, Francis and Hanson, Kael and Hebecker, Dustin and Heereman, David and Helbing, Klaus and Hellauer, Robert and Hickford, Stephanie and Hignight, Joshua and Hill, Gary and Hoffman, Kara and Hoffmann, Ruth and Hoinka, Tobias and Hokanson-Fasig, Benjamin and Hoshina, Kotoyo and Huang, Feifei and Huber, Matthias and Hultqvist, Klas and Huennefeld, Mirco and Hussain, Raamis and In, Seongjin and Iovine, Nadège and Ishihara, Aya and Jacobi, Emanuel and Japaridze, George and Jeong, Minjin and Jero, Kyle and Jones, Benjamin and Kalaczynski, Piotr and Kang, Woosik and Kappes, Alexander and Kappesser, David and Karg, Timo and Karle, Albrecht and Katz, Uli and Kauer, Matt and Keivani, Azadeh and Kelley, John and Kheirandish, Ali and Kim, JongHyun and Kim, Myoungchul and Kintscher, Thomas and Kiryluk, Joanna and Kittler, Thomas and Klein, Spencer and Koirala, Ramesh and Kolanoski, Hermann and Koepke, Lutz and Kopper, Claudio and Kopper, Sandro and Koschinsky, Jan Paul and Koskinen, Jason and Kowalski, Marek and Krammer, Benedikt and Krings, Kai and Kroll, Mike and Krueckl, Gerald and Kunwar, Samridha and Neilson, Naoko Kurahashi and Kuwabara, Takao and Kyriacou, Alexander and Labare, Mathieu and Lanfranchi, Justin and Larson, Michael and Lauber, Frederik and Leonard, Kayla and Lesiak-Bzdak, Mariola and Leuermann, Martin and Liu, Qinrui and Lozano Mariscal, Cristian Jesús and Lu, Lu and Luenemann, Jan and Luszczak, William and Madsen, James and Maggi, Giuliano and Mahn, Kendall and Mancina, Sarah and Maruyama, Reina and Mase, Keiichi and Maunu, Ryan and Meagher, Kevin and Medici, Morten and Meier, Maximilian and Menne, Thorben and Merino, Gonzalo and Meures, Thomas and Miarecki, Sandy and Micallef, Jessie and Momente, Giulio and Montaruli, Teresa and Moore, Roger and Morse, Robert and Moulai, Marjon and Nahnhauer, Rolf and Nakarmi, Prabandha and Naumann, Uwe and Neer, Garrett and Niederhausen, Hans and Nowicki, Sarah and Nygren, Dave and Pollmann, Anna and Olivas, Alex and Ó Murchadha, Aongus and O’Sullivan, Erin and Padovani, Paolo and Palczewski, Tomasz and Pandya, Hershal and Pankova, Daria and Peiffer, Peter and Pepper, James and Perez de los Heros, Carlos and Pieloth, Damian and Pinat, Elisa and Plum, Matthias and Price, Buford and Przybylski, Gerald and Raab, Christoph and Raedel, Leif and Rameez, Mohamed and Rawlins, Katherine and Rea, Immacolata Carmen and Reimann, Rene and Relethford, Ben and Relich, Matt and Resconi, Elisa and Rhode, Wolfgang and Richman, Mike and Robertson, Sally and Rongen, Martin and Rott, Carsten and Ruhe, Tim and Ryckbosch, Dirk and Rysewyk, Devyn and Safa, Ibrahim and Saelzer, Tobias and Sahakyan, Narek and Sanchez Herrera, Sebastian and Sandrock, Alexander and Sandroos, Joakim and Santander, Marcos and Sarkar, Sourav and Sarkar, Subir and Satalecka, Konstancja and Schlunder, Philipp and Schmidt, Torsten and Schneider, Austin and Schoenen, Sebastian and Schoeneberg, Sebastian and Schumacher, Lisa and Sclanfani, Stephen and Seckel, Dave and Seunarine, Suruj and Soedingrekso, Jan and Soldin, Dennis and Song, Ming and Spiczak, Glenn and Spiering, Christian and Stachurska, Juliana and Stamatikos, Michael and Stanev, Todor and Stasik, Alexander and Stettner, Joeran and Steuer, Anna and Stezelberger, Thorsten and Stokstad, Robert and Stoessl, Achim and Strotjohann, Nora Linn and Stuttard, Thomas and Sullivan, Greg and Sutherland, Michael and Taboada, Ignacio and Tatar, Joulien and Tenholt, Frederik and Ter-Antonyan, Samvel and Terliuk, Andrii and Tilav, Serap and Toale, Pat and Tobin, Moriah and Toennis, Christoph and Toscano, Simona and Tosi, Delia and Tselengidou, Maria and Tung, ChunFai and Turcati, Andrea and Turley, Colin and Ty, Bunheng and Unger, Lisa and Usner, Marcel and Van Driessche, Ward and Van Eijk, Daan and van Eijndhoven, Nick and Vandenbroucke, Justin and Vanheule, Sander and van Santen, Jakob and Vogel, Eric and Vraeghe, Matthias and Walck, Christian and Wallace, Alexander and Wallraff, Marius and Wandler, Frank and Wandkowsky, Nancy and Waza, Aatif and Weaver, Chris and Weiss, Matthew and Wendt, Chris and Werthebach, Johannes and Westerhoff, Stefan and Whelan, Ben and Whitehorn, Nathan and Wiebe, Klaus and Wiebusch, Christopher and Wille, Logan and Williams, Dawn and Wills, Lizz and Wolf, Martin and Wood, Joshua and Wood, Tania and Woschnagg, Kurt and Xu, Donglian and Xu, Xianwu and Xu, Yiqian and Yanez, Juan Pablo and Yodh, Gaurang and Yoshida, Shigeru and Yuan, Tianlu},
  journal   = {Science},
  title     = {Neutrino emission from the direction of the blazar TXS 0506+056 prior to the IceCube-170922A alert},
  year      = {2018},
  issn      = {1095-9203},
  month     = jul,
  number    = {6398},
  pages     = {147--151},
  volume    = {361},
  doi       = {10.1126/science.aat2890},
  publisher = {American Association for the Advancement of Science (AAAS)},
}

@Article{Albert2007,
  author    = {Albert, J. and Aliu, E. and Anderhub, H. and Antoranz, P. and Armada, A. and Baixeras, C. and Barrio, J. A. and Bartko, H. and Bastieri, D. and Becker, J. K. and Bednarek, W. and Berger, K. and Bigongiari, C. and Biland, A. and Bock, R. K. and Bordas, P. and Bosch‐Ramon, V. and Bretz, T. and Britvitch, I. and Camara, M. and Carmona, E. and Chilingarian, A. and Coarasa, J. A. and Commichau, S. and Contreras, J. L. and Cortina, J. and Costado, M. T. and Curtef, V. and Danielyan, V. and Dazzi, F. and De Angelis, A. and Delgado, C. and de los Reyes, R. and De Lotto, B. and Domingo‐Santamaria, E. and Dorner, D. and Doro, M. and Errando, M. and Fagiolini, M. and Ferenc, D. and Fernandez, E. and Firpo, R. and Flix, J. and Fonseca, M. V. and Font, L. and Fuchs, M. and Galante, N. and Garcia‐Lopez, R. J. and Garczarczyk, M. and Gaug, M. and Giller, M. and Goebel, F. and Hakobyan, D. and Hayashida, M. and Hengstebeck, T. and Herrero, A. and Hohne, D. and Hose, J. and Hrupec, D. and Hsu, C. C. and Jacon, P. and Jogler, T. and Kosyra, R. and Kranich, D. and Kritzer, R. and Laille, A. and Lindfors, E. and Lombardi, S. and Longo, F. and Lopez, J. and Lopez, M. and Lorenz, E. and Majumdar, P. and Maneva, G. and Mannheim, K. and Mansutti, O. and Mariotti, M. and Martinez, M. and Mazin, D. and Merck, C. and Meucci, M. and Meyer, M. and Miranda, J. M. and Mirzoyan, R. and Mizobuchi, S. and Moralejo, A. and Nieto, D. and Nilsson, K. and Ninkovic, J. and Ona‐Wilhelmi, E. and Otte, N. and Oya, I. and Paneque, D. and Panniello, M. and Paoletti, R. and Paredes, J. M. and Pasanen, M. and Pascoli, D. and Pauss, F. and Pegna, R. and Persic, M. and Peruzzo, L. and Piccioli, A. and Prandini, E. and Puchades, N. and Raymers, A. and Rhode, W. and Ribo, M. and Rico, J. and Rissi, M. and Robert, A. and Rugamer, S. and Saggion, A. and Saito, T. and Sanchez, A. and Sartori, P. and Scalzotto, V. and Scapin, V. and Schmitt, R. and Schweizer, T. and Shayduk, M. and Shinozaki, K. and Shore, S. N. and Sidro, N. and Sillanpaa, A. and Sobczynska, D. and Stamerra, A. and Stark, L. S. and Takalo, L. and Tavecchio, F. and Temnikov, P. and Tescaro, D. and Teshima, M. and Torres, D. F. and Turini, N. and Vankov, H. and Vitale, V. and Wagner, R. M. and Wibig, T. and Wittek, W. and Zandanel, F. and Zanin, R. and Zapatero, J.},
  journal   = {The Astrophysical Journal},
  title     = {Variable Very High Energy γ‐Ray Emission from Markarian 501},
  year      = {2007},
  issn      = {1538-4357},
  month     = nov,
  number    = {2},
  pages     = {862--883},
  volume    = {669},
  doi       = {10.1086/521382},
  publisher = {American Astronomical Society},
}

@Article{Ackermann2016,
  author    = {Ackermann, M. and Anantua, R. and Asano, K. and Baldini, L. and Barbiellini, G. and Bastieri, D. and Gonzalez, J. Becerra and Bellazzini, R. and Bissaldi, E. and Blandford, R. D. and Bloom, E. D. and Bonino, R. and Bottacini, E. and Bruel, P. and Buehler, R. and Caliandro, G. A. and Cameron, R. A. and Caragiulo, M. and Caraveo, P. A. and Cavazzuti, E. and Cecchi, C. and Cheung, C. C. and Chiang, J. and Chiaro, G. and Ciprini, S. and Cohen-Tanugi, J. and Costanza, F. and Cutini, S. and D’Ammando, F. and Palma, F. de and Desiante, R. and Digel, S. W. and Lalla, N. Di and Mauro, M. Di and Venere, L. Di and Drell, P. S. and Favuzzi, C. and Fegan, S. J. and Ferrara, E. C. and Fukazawa, Y. and Funk, S. and Fusco, P. and Gargano, F. and Gasparrini, D. and Giglietto, N. and Giordano, F. and Giroletti, M. and Grenier, I. A. and Guillemot, L. and Guiriec, S. and Hayashida, M. and Hays, E. and Horan, D. and Jóhannesson, G. and Kensei, S. and Kocevski, D. and Kuss, M. and Mura, G. La and Larsson, S. and Latronico, L. and Li, J. and Longo, F. and Loparco, F. and Lott, B. and Lovellette, M. N. and Lubrano, P. and Madejski, G. M. and Magill, J. D. and Maldera, S. and Manfreda, A. and Mayer, M. and Mazziotta, M. N. and Michelson, P. F. and Mirabal, N. and Mizuno, T. and Monzani, M. E. and Morselli, A. and Moskalenko, I. V. and Nalewajko, K. and Negro, M. and Nuss, E. and Ohsugi, T. and Orlando, E. and Paneque, D. and Perkins, J. S. and Pesce-Rollins, M. and Piron, F. and Pivato, G. and Porter, T. A. and Principe, G. and Rando, R. and Razzano, M. and Razzaque, S. and Reimer, A. and Scargle, J. D. and Sgrò, C. and Sikora, M. and Simone, D. and Siskind, E. J. and Spada, F. and Spinelli, P. and Stawarz, L. and Thayer, J. B. and Thompson, D. J. and Torres, D. F. and Troja, E. and Uchiyama, Y. and Yuan, Y. and Zimmer, S.},
  journal   = {The Astrophysical Journal Letters},
  title     = {MINUTE-TIMESCALE $>$100 MeV γ-RAY VARIABILITY DURING THE GIANT OUTBURST OF QUASAR 3C 279 OBSERVED BY FERMI-LAT IN 2015 JUNE},
  year      = {2016},
  issn      = {2041-8213},
  month     = jun,
  number    = {2},
  pages     = {L20},
  volume    = {824},
  doi       = {10.3847/2041-8205/824/2/l20},
  publisher = {American Astronomical Society},
}

@Article{Shukla2020,
  author    = {Shukla, A. and Mannheim, K.},
  journal   = {Nature Communications},
  title     = {Gamma-ray flares from relativistic magnetic reconnection in the jet of the quasar 3C 279},
  year      = {2020},
  issn      = {2041-1723},
  month     = aug,
  number    = {1},
  volume    = {11},
  doi       = {10.1038/s41467-020-17912-z},
  publisher = {Springer Science and Business Media LLC},
}

@Article{Mignone2007,
  author    = {Mignone, A. and Bodo, G. and Massaglia, S. and Matsakos, T. and Tesileanu, O. and Zanni, C. and Ferrari, A.},
  journal   = {The Astrophysical Journal Supplement Series},
  title     = {PLUTO: A Numerical Code for Computational Astrophysics},
  year      = {2007},
  issn      = {1538-4365},
  month     = may,
  number    = {1},
  pages     = {228--242},
  volume    = {170},
  doi       = {10.1086/513316},
  publisher = {American Astronomical Society},
}

@Article{Mignone2019,
  author    = {Mignone, A and Mattia, G and Bodo, G and Del Zanna, L},
  journal   = {Monthly Notices of the Royal Astronomical Society},
  title     = {A constrained transport method for the solution of the resistive relativistic MHD equations},
  year      = {2019},
  issn      = {1365-2966},
  month     = apr,
  number    = {3},
  pages     = {4252--4274},
  volume    = {486},
  doi       = {10.1093/mnras/stz1015},
  publisher = {Oxford University Press (OUP)},
}

@Article{Mignone2007a,
  author    = {Mignone, A. and McKinney, Jonathan C.},
  journal   = {Monthly Notices of the Royal Astronomical Society},
  title     = {Equation of state in relativistic magnetohydrodynamics: variable versus constant adiabatic index},
  year      = {2007},
  issn      = {1365-2966},
  month     = jun,
  number    = {3},
  pages     = {1118--1130},
  volume    = {378},
  doi       = {10.1111/j.1365-2966.2007.11849.x},
  publisher = {Oxford University Press (OUP)},
}

@Article{Striani2016,
  author    = {Striani, E. and Mignone, A. and Vaidya, B. and Bodo, G. and Ferrari, A.},
  journal   = {Monthly Notices of the Royal Astronomical Society},
  title     = {MHD simulations of three-dimensional resistive reconnection in a cylindrical plasma column},
  year      = {2016},
  issn      = {1365-2966},
  month     = jul,
  number    = {3},
  pages     = {2970--2979},
  volume    = {462},
  doi       = {10.1093/mnras/stw1848},
  publisher = {Oxford University Press (OUP)},
}

@Article{Bodo2019,
  author    = {Bodo, G and Mamatsashvili, G and Rossi, P and Mignone, A},
  journal   = {Monthly Notices of the Royal Astronomical Society},
  title     = {Linear stability analysis of magnetized relativistic rotating jets},
  year      = {2019},
  issn      = {1365-2966},
  month     = mar,
  number    = {2},
  pages     = {2909--2921},
  volume    = {485},
  doi       = {10.1093/mnras/stz591},
  publisher = {Oxford University Press (OUP)},
}

@Article{Bromberg2019,
  author    = {Bromberg, Omer and Singh, Chandra B. and Davelaar, Jordy and Philippov, Alexander A.},
  journal   = {The Astrophysical Journal},
  title     = {Kink Instability: Evolution and Energy Dissipation in Relativistic Force-free Nonrotating Jets},
  year      = {2019},
  issn      = {1538-4357},
  month     = oct,
  number    = {1},
  pages     = {39},
  volume    = {884},
  doi       = {10.3847/1538-4357/ab3fa5},
  publisher = {American Astronomical Society},
}

@Article{Bodo2021,
  author    = {Bodo, G and Mamatsashvili, G and Rossi, P and Mignone, A},
  journal   = {Monthly Notices of the Royal Astronomical Society},
  title     = {Current-driven kink instabilities in relativistic jets: dissipation properties},
  year      = {2021},
  issn      = {1365-2966},
  month     = dec,
  number    = {2},
  pages     = {2391--2406},
  volume    = {510},
  doi       = {10.1093/mnras/stab3492},
  publisher = {Oxford University Press (OUP)},
}

@Article{Bodo2020,
  author    = {Bodo, G and Tavecchio, F and Sironi, L},
  journal   = {Monthly Notices of the Royal Astronomical Society},
  title     = {Kink-driven magnetic reconnection in relativistic jets: consequences for X-ray polarimetry of BL Lacs},
  year      = {2020},
  issn      = {1365-2966},
  month     = nov,
  number    = {2},
  pages     = {2836--2847},
  volume    = {501},
  doi       = {10.1093/mnras/staa3620},
  publisher = {Oxford University Press (OUP)},
}

@Article{Hesse1988,
  author    = {Hesse, M. and Schindler, K.},
  journal   = {Journal of Geophysical Research: Space Physics},
  title     = {A theoretical foundation of general magnetic reconnection},
  year      = {1988},
  issn      = {0148-0227},
  month     = jun,
  number    = {A6},
  pages     = {5559--5567},
  volume    = {93},
  doi       = {10.1029/ja093ia06p05559},
  publisher = {American Geophysical Union (AGU)},
}

@Article{Houlahan1992,
  author    = {Houlahan, Padraig and Scalo, John},
  journal   = {The Astrophysical Journal},
  title     = {Recognition and characterization of hierarchical interstellar structure. II - Structure tree statistics},
  year      = {1992},
  issn      = {1538-4357},
  month     = jul,
  pages     = {172},
  volume    = {393},
  doi       = {10.1086/171495},
  publisher = {American Astronomical Society},
}

@Article{Rosolowsky2008,
  author    = {Rosolowsky, E. W. and Pineda, J. E. and Kauffmann, J. and Goodman, A. A.},
  journal   = {The Astrophysical Journal},
  title     = {Structural Analysis of Molecular Clouds: Dendrograms},
  year      = {2008},
  issn      = {1538-4357},
  month     = jun,
  number    = {2},
  pages     = {1338--1351},
  volume    = {679},
  doi       = {10.1086/587685},
  publisher = {American Astronomical Society},
}

@Article{Piperno2020,
  author    = {Piperno, Enrico and Guszejnov, Dávid and Offner, Stella S. R. and Grudić, Michael Y.},
  journal   = {Research Notes of the AAS},
  title     = {Comparing Methods to Identify GMCs in Simulated Galaxies},
  year      = {2020},
  issn      = {2515-5172},
  month     = jan,
  number    = {1},
  pages     = {14},
  volume    = {4},
  doi       = {10.3847/2515-5172/ab7022},
  publisher = {American Astronomical Society},
}

@Article{Kadowaki2021,
  author    = {Kadowaki, Luis H. S. and de Gouveia Dal Pino, Elisabete M. and Medina-Torrejón, Tania E. and Mizuno, Yosuke and Kushwaha, Pankaj},
  journal   = {The Astrophysical Journal},
  title     = {Fast Magnetic Reconnection Structures in Poynting Flux-dominated Jets},
  year      = {2021},
  issn      = {1538-4357},
  month     = may,
  number    = {2},
  pages     = {109},
  volume    = {912},
  doi       = {10.3847/1538-4357/abee7a},
  publisher = {American Astronomical Society},
}

@Article{Rossi2008,
  author    = {Rossi, P. and Mignone, A. and Bodo, G. and Massaglia, S. and Ferrari, A.},
  journal   = {Astronomy \& Astrophysics},
  title     = {Formation of dynamical structures in relativistic jets: the FRI case},
  year      = {2008},
  issn      = {1432-0746},
  month     = jul,
  number    = {3},
  pages     = {795--806},
  volume    = {488},
  doi       = {10.1051/0004-6361:200809687},
  publisher = {EDP Sciences},
}

@Article{Ghisellini1996,
  author    = {Ghisellini, G. and Madau, P.},
  journal   = {Monthly Notices of the Royal Astronomical Society},
  title     = {On the origin of the  -ray emission in blazars},
  year      = {1996},
  issn      = {1365-2966},
  month     = may,
  number    = {1},
  pages     = {67--76},
  volume    = {280},
  doi       = {10.1093/mnras/280.1.67},
  publisher = {Oxford University Press (OUP)},
}

@Article{Liu2006,
  author    = {Liu, H. T. and Bai, J. M.},
  journal   = {The Astrophysical Journal},
  title     = {Absorption of 10–200 GeV Gamma Rays by Radiation from Broad‐Line Regions in Blazars},
  year      = {2006},
  issn      = {1538-4357},
  month     = dec,
  number    = {2},
  pages     = {1089--1097},
  volume    = {653},
  doi       = {10.1086/509097},
  publisher = {American Astronomical Society},
}

@Article{Ghisellini2009,
  author    = {Ghisellini, G. and Tavecchio, F.},
  journal   = {Monthly Notices of the Royal Astronomical Society},
  title     = {Canonical high-power blazars},
  year      = {2009},
  issn      = {1365-2966},
  month     = aug,
  number    = {2},
  pages     = {985--1002},
  volume    = {397},
  doi       = {10.1111/j.1365-2966.2009.15007.x},
  publisher = {Oxford University Press (OUP)},
}

@Article{Zhdankin2013,
  author    = {Zhdankin, Vladimir and Uzdensky, Dmitri A. and Perez, Jean C. and Boldyrev, Stanislav},
  journal   = {The Astrophysical Journal},
  title     = {STATISTICAL ANALYSIS OF CURRENT SHEETS IN THREE-DIMENSIONAL MAGNETOHYDRODYNAMIC TURBULENCE},
  year      = {2013},
  issn      = {1538-4357},
  month     = jun,
  number    = {2},
  pages     = {124},
  volume    = {771},
  doi       = {10.1088/0004-637x/771/2/124},
  publisher = {American Astronomical Society},
}

@Article{Nurisso2023,
  author    = {Nurisso, Matteo and Celotti, Annalisa and Mignone, Andrea and Bodo, Gianluigi},
  journal   = {Monthly Notices of the Royal Astronomical Society},
  title     = {Particle acceleration with magnetic reconnection in large-scale RMHD simulations – I. Current sheet identification and characterization},
  year      = {2023},
  issn      = {1365-2966},
  month     = may,
  number    = {4},
  pages     = {5517--5528},
  volume    = {522},
  doi       = {10.1093/mnras/stad1348},
  publisher = {Oxford University Press (OUP)},
}

@misc{Robitaille2019,
  author       = {Robitaille, Thomas and Rice, Tom and Beaumont, Chris and Ginsburg, Adam and MacDonald, Braden and Rosolowsky, Erik},
  eid          = {ascl:1907.016},
  howpublished = {Astrophysics Source Code Library, record ascl:1907.016},
  month        = jul,
  title        = {astrodendro: Astronomical data dendrogram creator},
  url          = {https://ui.adsabs.harvard.edu/abs/2019ascl.soft07016R},
  year         = {2019},
}

@Article{Johnston2014,
  author    = {Johnston, K. G. and Beuther, H. and Linz, H. and Schmiedeke, A. and Ragan, S. E. and Henning, Th.},
  journal   = {Astronomy \& Astrophysics},
  title     = {The dynamics and star-forming potential of the massive Galactic centre cloud G0.253+0.016},
  year      = {2014},
  issn      = {1432-0746},
  month     = aug,
  pages     = {A56},
  volume    = {568},
  doi       = {10.1051/0004-6361/201423943},
  publisher = {EDP Sciences},
}

@misc{Wang2024a,
  author       = {Wang, Yulei and Cheng, Xin and Guo, Yang and Guo, Jinhan and Ding, Mingde},
  eid          = {ascl:2401.014},
  howpublished = {Astrophysics Source Code Library, record ascl:2401.014},
  month        = jan,
  title        = {LoRD: Locate Reconnection Distribution},
  url          = {https://ui.adsabs.harvard.edu/abs/2024ascl.soft01014W},
  year         = {2024},
}

@Article{Giannios2009,
  author    = {Giannios, Dimitrios and Uzdensky, Dmitri A. and Begelman, Mitchell C.},
  journal   = {Monthly Notices of the Royal Astronomical Society: Letters},
  title     = {Fast TeV variability in blazars: jets in a jet},
  year      = {2009},
  issn      = {1745-3925},
  month     = may,
  number    = {1},
  pages     = {L29--L33},
  volume    = {395},
  doi       = {10.1111/j.1745-3933.2009.00635.x},
  publisher = {Oxford University Press (OUP)},
}

@Article{Dubey2023,
  author    = {Dubey, Ravi Pratap and Fendt, Christian and Vaidya, Bhargav},
  journal   = {The Astrophysical Journal},
  title     = {Particles in Relativistic MHD Jets. I. Role of Jet Dynamics in Particle Acceleration},
  year      = {2023},
  issn      = {1538-4357},
  month     = jul,
  number    = {1},
  pages     = {1},
  volume    = {952},
  doi       = {10.3847/1538-4357/ace0bf},
  publisher = {American Astronomical Society},
}

@Article{Abramowski2010,
  author    = {Abramowski, A. and Acero, F. and Aharonian, F. and Akhperjanian, A. G. and Anton, G. and Barres de Almeida, U. and Bazer-Bachi, A. R. and Becherini, Y. and Behera, B. and Benbow, W. and Bernlöhr, K. and Bochow, A. and Boisson, C. and Bolmont, J. and Borrel, V. and Brucker, J. and Brun, F. and Brun, P. and Bühler, R. and Bulik, T. and Büsching, I. and Boutelier, T. and Chadwick, P. M. and Charbonnier, A. and Chaves, R. C. G. and Cheesebrough, A. and Chounet, L.-M. and Clapson, A. C. and Coignet, G. and Conrad, J. and Costamante, L. and Dalton, M. and Daniel, M. K. and Davids, I. D. and Degrange, B. and Deil, C. and Dickinson, H. J. and Djannati-Ataï, A. and Domainko, W. and Drury, L. O’C. and Dubois, F. and Dubus, G. and Dyks, J. and Dyrda, M. and Egberts, K. and Eger, P. and Espigat, P. and Fallon, L. and Farnier, C. and Fegan, S. and Feinstein, F. and Fernandes, M. V. and Fiasson, A. and Förster, A. and Fontaine, G. and Füßling, M. and Gabici, S. and Gallant, Y. A. and Gérard, L. and Gerbig, D. and Giebels, B. and Glicenstein, J. F. and Glück, B. and Goret, P. and Göring, D. and Hampf, D. and Hauser, M. and Heinz, S. and Heinzelmann, G. and Henri, G. and Hermann, G. and Hinton, J. A. and Hoffmann, A. and Hofmann, W. and Hofverberg, P. and Holleran, M. and Hoppe, S. and Horns, D. and Jacholkowska, A. and de Jager, O. C. and Jahn, C. and Jung, I. and Katarzyński, K. and Katz, U. and Kaufmann, S. and Kerschhaggl, M. and Khangulyan, D. and Khélifi, B. and Keogh, D. and Klochkov, D. and Kluźniak, W. and Kneiske, T. and Komin, Nu. and Kosack, K. and Kossakowski, R. and Lamanna, G. and Lenain, J.-P. and Lohse, T. and Lu, C.-C. and Marandon, V. and Marcowith, A. and Masbou, J. and Maurin, D. and McComb, T. J. L. and Medina, M. C. and Méhault, J. and Moderski, R. and Moulin, E. and Naumann-Godo, M. and de Naurois, M. and Nedbal, D. and Nekrassov, D. and Nguyen, N. and Nicholas, B. and Niemiec, J. and Nolan, S. J. and Ohm, S. and Olive, J.-F. and de Oña Wilhelmi, E. and Opitz, B. and Orford, K. J. and Ostrowski, M. and Panter, M. and Paz Arribas, M. and Pedaletti, G. and Pelletier, G. and Petrucci, P.-O. and Pita, S. and Pühlhofer, G. and Punch, M. and Quirrenbach, A. and Raubenheimer, B. C. and Raue, M. and Rayner, S. M. and Reimer, O. and Renaud, M. and de los Reyes, R. and Rieger, F. and Ripken, J. and Rob, L. and Rosier-Lees, S. and Rowell, G. and Rudak, B. and Rulten, C. B. and Ruppel, J. and Ryde, F. and Sahakian, V. and Santangelo, A. and Schlickeiser, R. and Schöck, F. M. and Schönwald, A. and Schwanke, U. and Schwarzburg, S. and Schwemmer, S. and Shalchi, A. and Sushch, I. and Sikora, M. and Skilton, J. L. and Sol, H. and Stawarz, Ł. and Steenkamp, R. and Stegmann, C. and Stinzing, F. and Superina, G. and Szostek, A. and Tam, P. H. and Tavernet, J.-P. and Terrier, R. and Tibolla, O. and Tluczykont, M. and Valerius, K. and van Eldik, C. and Vasileiadis, G. and Venter, C. and Venter, L. and Vialle, J. P. and Viana, A. and Vincent, P. and Vivier, M. and Völk, H. J. and Volpe, F. and Vorobiov, S. and Wagner, S. J. and Ward, M. and Zdziarski, A. A. and Zech, A. and Zechlin, H.-S.},
  journal   = {Astronomy and Astrophysics},
  title     = {VHEγ-ray emission of PKS 2155–304: spectral and temporal variability},
  year      = {2010},
  issn      = {1432-0746},
  month     = sep,
  pages     = {A83},
  volume    = {520},
  doi       = {10.1051/0004-6361/201014484},
  publisher = {EDP Sciences},
}

@Article{Zhdankin2017,
  author    = {Zhdankin, Vladimir and Werner, Gregory R. and Uzdensky, Dmitri A. and Begelman, Mitchell C.},
  journal   = {Physical Review Letters},
  title     = {Kinetic Turbulence in Relativistic Plasma: From Thermal Bath to Nonthermal Continuum},
  year      = {2017},
  issn      = {1079-7114},
  month     = feb,
  number    = {5},
  pages     = {055103},
  volume    = {118},
  doi       = {10.1103/physrevlett.118.055103},
  publisher = {American Physical Society (APS)},
}

@Article{Lazarian1999,
  author    = {Lazarian, A. and Vishniac, Ethan T.},
  journal   = {The Astrophysical Journal},
  title     = {Reconnection in a Weakly Stochastic Field},
  year      = {1999},
  issn      = {1538-4357},
  month     = jun,
  number    = {2},
  pages     = {700--718},
  volume    = {517},
  doi       = {10.1086/307233},
  publisher = {American Astronomical Society},
}

@Article{Blandford2019,
  author    = {Blandford, Roger and Meier, David and Readhead, Anthony},
  journal   = {Annual Review of Astronomy and Astrophysics},
  title     = {Relativistic Jets from Active Galactic Nuclei},
  year      = {2019},
  issn      = {1545-4282},
  month     = aug,
  number    = {1},
  pages     = {467--509},
  volume    = {57},
  doi       = {10.1146/annurev-astro-081817-051948},
  publisher = {Annual Reviews},
}

@Article{Kim2021,
  author    = {Kim, Dohyeong and Lee, Daye and Im, Myungshin},
  journal   = {Monthly Notices of the Royal Astronomical Society},
  title     = {Bolometric luminosity estimators using infrared hydrogen lines for dust obscured active galactic nuclei},
  year      = {2021},
  issn      = {1365-2966},
  month     = oct,
  number    = {1},
  pages     = {1147--1159},
  volume    = {509},
  doi       = {10.1093/mnras/stab3072},
  publisher = {Oxford University Press (OUP)},
}

@Article{Blandford1987,
  author    = {Blandford, Roger and Eichler, David},
  journal   = {Physics Reports},
  title     = {Particle acceleration at astrophysical shocks: A theory of cosmic ray origin},
  year      = {1987},
  issn      = {0370-1573},
  month     = oct,
  number    = {1},
  pages     = {1--75},
  volume    = {154},
  doi       = {10.1016/0370-1573(87)90134-7},
  publisher = {Elsevier BV},
}

@Article{Barkov2012,
  author    = {Barkov, M. V. and Aharonian, F. A. and Bogovalov, S. V. and Kelner, S. R. and Khangulyan, D.},
  journal   = {The Astrophysical Journal},
  title     = {RAPID TeV VARIABILITY IN BLAZARS AS A RESULT OF JET-STAR INTERACTION},
  year      = {2012},
  issn      = {1538-4357},
  month     = mar,
  number    = {2},
  pages     = {119},
  volume    = {749},
  doi       = {10.1088/0004-637x/749/2/119},
  publisher = {American Astronomical Society},
}

@Article{Mattia2023,
  author    = {Mattia, G. and Del Zanna, L. and Bugli, M. and Pavan, A. and Ciolfi, R. and Bodo, G. and Mignone, A.},
  journal   = {Astronomy \& Astrophysics},
  title     = {Resistive relativistic MHD simulations of astrophysical jets},
  year      = {2023},
  issn      = {1432-0746},
  month     = nov,
  pages     = {A49},
  volume    = {679},
  doi       = {10.1051/0004-6361/202347126},
  publisher = {EDP Sciences},
}

@Article{Urry1995,
  author    = {Urry, C. Megan and Padovani, Paolo},
  journal   = {Publications of the Astronomical Society of the Pacific},
  title     = {Unified Schemes for Radio-Loud Active Galactic Nuclei},
  year      = {1995},
  issn      = {1538-3873},
  month     = sep,
  pages     = {803},
  volume    = {107},
  doi       = {10.1086/133630},
  publisher = {IOP Publishing},
}

@Article{Sikora2007,
  author    = {Sikora, Marek and Stawarz, Łukasz and Lasota, Jean‐Pierre},
  journal   = {The Astrophysical Journal},
  title     = {Radio Loudness of Active Galactic Nuclei: Observational Facts and Theoretical Implications},
  year      = {2007},
  issn      = {1538-4357},
  month     = apr,
  number    = {2},
  pages     = {815--828},
  volume    = {658},
  doi       = {10.1086/511972},
  publisher = {American Astronomical Society},
}

@Article{Puzzoni2021,
  author    = {Puzzoni, E and Mignone, A and Bodo, G},
  journal   = {Monthly Notices of the Royal Astronomical Society},
  title     = {On the impact of the numerical method on magnetic reconnection and particle acceleration – I. The MHD case},
  year      = {2021},
  issn      = {1365-2966},
  month     = oct,
  number    = {2},
  pages     = {2771--2783},
  volume    = {508},
  doi       = {10.1093/mnras/stab2813},
  publisher = {Oxford University Press (OUP)},
}

@Article{BarniolDuran2017,
  author    = {Barniol Duran, Rodolfo and Tchekhovskoy, Alexander and Giannios, Dimitrios},
  journal   = {Monthly Notices of the Royal Astronomical Society},
  title     = {Simulations of AGN jets: magnetic kink instability versus conical shocks},
  year      = {2017},
  issn      = {1365-2966},
  month     = may,
  number    = {4},
  pages     = {4957--4978},
  volume    = {469},
  doi       = {10.1093/mnras/stx1165},
  publisher = {Oxford University Press (OUP)},
}

@Article{Petropoulou2019,
  author    = {Petropoulou, Maria and Sironi, Lorenzo and Spitkovsky, Anatoly and Giannios, Dimitrios},
  journal   = {The Astrophysical Journal},
  title     = {Relativistic Magnetic Reconnection in Electron–Positron–Proton Plasmas: Implications for Jets of Active Galactic Nuclei},
  year      = {2019},
  issn      = {1538-4357},
  month     = jul,
  number    = {1},
  pages     = {37},
  volume    = {880},
  doi       = {10.3847/1538-4357/ab287a},
  publisher = {American Astronomical Society},
}

@Article{Kadowaki2018,
  author    = {Kadowaki, Luis H. S. and De Gouveia Dal Pino, Elisabete M. and Stone, James M.},
  journal   = {The Astrophysical Journal},
  title     = {MHD Instabilities in Accretion Disks and Their Implications in Driving Fast Magnetic Reconnection},
  year      = {2018},
  issn      = {1538-4357},
  month     = aug,
  number    = {1},
  pages     = {52},
  volume    = {864},
  doi       = {10.3847/1538-4357/aad4ff},
  publisher = {American Astronomical Society},
}

@Article{Aleksic2011,
  author    = {Aleksić, J. and Antonelli, L. A. and Antoranz, P. and Backes, M. and Barrio, J. A. and Bastieri, D. and Becerra González, J. and Bednarek, W. and Berdyugin, A. and Berger, K. and Bernardini, E. and Biland, A. and Blanch, O. and Bock, R. K. and Boller, A. and Bonnoli, G. and Borla Tridon, D. and Braun, I. and Bretz, T. and Cañellas, A. and Carmona, E. and Carosi, A. and Colin, P. and Colombo, E. and Contreras, J. L. and Cortina, J. and Cossio, L. and Covino, S. and Dazzi, F. and De Angelis, A. and De Cea del Pozo, E. and De Lotto, B. and Delgado Mendez, C. and Diago Ortega, A. and Doert, M. and Domínguez, A. and Dominis Prester, D. and Dorner, D. and Doro, M. and Elsaesser, D. and Ferenc, D. and Fonseca, M. V. and Font, L. and Fruck, C. and García López, R. J. and Garczarczyk, M. and Garrido, D. and Giavitto, G. and Godinović, N. and Hadasch, D. and Häfner, D. and Herrero, A. and Hildebrand, D. and Höhne-Mönch, D. and Hose, J. and Hrupec, D. and Huber, B. and Jogler, T. and Klepser, S. and Krähenbühl, T. and Krause, J. and La Barbera, A. and Lelas, D. and Leonardo, E. and Lindfors, E. and Lombardi, S. and López, M. and Lorenz, E. and Makariev, M. and Maneva, G. and Mankuzhiyil, N. and Mannheim, K. and Maraschi, L. and Mariotti, M. and Martínez, M. and Mazin, D. and Meucci, M. and Miranda, J. M. and Mirzoyan, R. and Miyamoto, H. and Moldón, J. and Moralejo, A. and Nieto, D. and Nilsson, K. and Orito, R. and Oya, I. and Paneque, D. and Paoletti, R. and Pardo, S. and Paredes, J. M. and Partini, S. and Pasanen, M. and Pauss, F. and Perez-Torres, M. A. and Persic, M. and Peruzzo, L. and Pilia, M. and Pochon, J. and Prada, F. and Prada Moroni, P. G. and Prandini, E. and Puljak, I. and Reichardt, I. and Reinthal, R. and Rhode, W. and Ribó, M. and Rico, J. and Rügamer, S. and Saggion, A. and Saito, K. and Saito, T. Y. and Salvati, M. and Satalecka, K. and Scalzotto, V. and Scapin, V. and Schultz, C. and Schweizer, T. and Shayduk, M. and Shore, S. N. and Sillanpää, A. and Sitarek, J. and Sobczynska, D. and Spanier, F. and Spiro, S. and Stamerra, A. and Steinke, B. and Storz, J. and Strah, N. and Surić, T. and Takalo, L. and Tavecchio, F. and Temnikov, P. and Terzić, T. and Tescaro, D. and Teshima, M. and Thom, M. and Tibolla, O. and Torres, D. F. and Treves, A. and Vankov, H. and Vogler, P. and Wagner, R. M. and Weitzel, Q. and Zabalza, V. and Zandanel, F. and Zanin, R. and Tanaka, Y. T. and Wood, D. L. and Buson, S.},
  journal   = {The Astrophysical Journal},
  title     = {MAGIC DISCOVERY OF VERY HIGH ENERGY EMISSION FROM THE FSRQ PKS 1222+21},
  year      = {2011},
  issn      = {2041-8213},
  month     = feb,
  number    = {1},
  pages     = {L8},
  volume    = {730},
  doi       = {10.1088/2041-8205/730/1/l8},
  publisher = {American Astronomical Society},
}

@Article{Begelman2008,
  author    = {Begelman, Mitchell C. and Fabian, Andrew C. and Rees, Martin J.},
  journal   = {Monthly Notices of the Royal Astronomical Society: Letters},
  title     = {Implications of very rapid TeV variability in blazars},
  year      = {2008},
  issn      = {1745-3925},
  month     = feb,
  number    = {1},
  pages     = {L19--L23},
  volume    = {384},
  doi       = {10.1111/j.1745-3933.2007.00413.x},
  publisher = {Oxford University Press (OUP)},
}

@Article{Giannios2013,
  author    = {Giannios, Dimitrios},
  journal   = {Monthly Notices of the Royal Astronomical Society},
  title     = {Reconnection-driven plasmoids in blazars: fast flares on a slow envelope},
  year      = {2013},
  issn      = {1365-2966},
  month     = feb,
  number    = {1},
  pages     = {355--363},
  volume    = {431},
  doi       = {10.1093/mnras/stt167},
  publisher = {Oxford University Press (OUP)},
}

@Article{Giannios2006,
  author    = {Giannios, D. and Spruit, H. C.},
  journal   = {Astronomy \& Astrophysics},
  title     = {The role of kink instability in Poynting-flux dominated jets},
  year      = {2006},
  issn      = {1432-0746},
  month     = apr,
  number    = {3},
  pages     = {887--898},
  volume    = {450},
  doi       = {10.1051/0004-6361:20054107},
  publisher = {EDP Sciences},
}

@Article{Paul2022,
  author    = {Paul, Arghyadeep and Vaidya, Bhargav and Strugarek, Antoine},
  journal   = {The Astrophysical Journal},
  title     = {A Volumetric Study of Flux Transfer Events at the Dayside Magnetopause},
  year      = {2022},
  issn      = {1538-4357},
  month     = oct,
  number    = {2},
  pages     = {130},
  volume    = {938},
  doi       = {10.3847/1538-4357/ac8eb5},
  publisher = {American Astronomical Society},
}

@Article{Blandford1982,
  author    = {Blandford, R. D. and Payne, D. G.},
  journal   = {Monthly Notices of the Royal Astronomical Society},
  title     = {Hydromagnetic flows from accretion discs and the production of radio jets},
  year      = {1982},
  issn      = {1365-2966},
  month     = aug,
  number    = {4},
  pages     = {883--903},
  volume    = {199},
  doi       = {10.1093/mnras/199.4.883},
  publisher = {Oxford University Press (OUP)},
}

@Article{Mizuno2012,
  author    = {Mizuno, Yosuke and Lyubarsky, Yuri and Nishikawa, Ken-Ichi and Hardee, Philip E.},
  journal   = {The Astrophysical Journal},
  title     = {THREE-DIMENSIONAL RELATIVISTIC MAGNETOHYDRODYNAMIC SIMULATIONS OF CURRENT-DRIVEN INSTABILITY. III. ROTATING RELATIVISTIC JETS},
  year      = {2012},
  issn      = {1538-4357},
  month     = aug,
  number    = {1},
  pages     = {16},
  volume    = {757},
  doi       = {10.1088/0004-637x/757/1/16},
  publisher = {American Astronomical Society},
}

@Article{Kowal2017,
  author    = {Kowal, Grzegorz and Falceta-Gonçalves, Diego A. and Lazarian, Alex and Vishniac, Ethan T.},
  journal   = {The Astrophysical Journal},
  title     = {Statistics of Reconnection-driven Turbulence},
  year      = {2017},
  issn      = {1538-4357},
  month     = mar,
  number    = {2},
  pages     = {91},
  volume    = {838},
  doi       = {10.3847/1538-4357/aa6001},
  publisher = {American Astronomical Society},
}

@InBook{GouveiaDalPino2014,
  author    = {de Gouveia Dal Pino, Elisabete M. and Kowal, Grzegorz},
  pages     = {373--398},
  publisher = {Springer Berlin Heidelberg},
  title     = {Particle Acceleration by Magnetic Reconnection},
  year      = {2014},
  isbn      = {9783662446256},
  month     = oct,
  booktitle = {Magnetic Fields in Diffuse Media},
  doi       = {10.1007/978-3-662-44625-6_13},
  issn      = {2214-7985},
}

@Article{Mattia2024,
  author    = {Mattia, Giancarlo and Del Zanna, Luca and Pavan, Andrea and Ciolfi, Riccardo},
  journal   = {Astronomy \& Astrophysics},
  title     = {Magnetic dissipation in short gamma-ray-burst jets: I. Resistive relativistic MHD evolution in a model environment},
  year      = {2024},
  issn      = {1432-0746},
  month     = oct,
  pages     = {A105},
  volume    = {691},
  doi       = {10.1051/0004-6361/202451528},
  publisher = {EDP Sciences},
}

@Misc{Mattia2025,
  author    = {Mattia, Giancarlo and Crocco, Daniele and Fuksman, David Melon and Bugli, Matteo and Berta, Vittoria and Puzzoni, Eleonora and Mignone, Andrea and Vaidya, Bhargav},
  title     = {PyPLUTO: a data analysis Python package for the PLUTO code},
  year      = {2025},
  copyright = {arXiv.org perpetual, non-exclusive license},
  doi       = {10.48550/ARXIV.2501.09748},
  publisher = {arXiv},
}

@Article{Agarwal2023,
  author    = {Agarwal, S and Banerjee, B and Shukla, A and Roy, J and Acharya, S and Vaidya, B and Chitnis, V R and Wagner, S M and Mannheim, K and Branchesi, M},
  journal   = {Monthly Notices of the Royal Astronomical Society: Letters},
  title     = {Flaring activity from magnetic reconnection in BL Lacertae},
  year      = {2023},
  issn      = {1745-3933},
  month     = feb,
  number    = {1},
  pages     = {L53--L58},
  volume    = {521},
  doi       = {10.1093/mnrasl/slad023},
  publisher = {Oxford University Press (OUP)},
}

@Article{Kadler2012,
  author    = {Kadler, M. and Eisenacher, D. and Ros, E. and Mannheim, K. and Elsässer, D. and Bach, U.},
  journal   = {Astronomy \& Astrophysics},
  title     = {The blazar-like radio structure of the TeV source IC 310},
  year      = {2012},
  issn      = {1432-0746},
  month     = jan,
  pages     = {L1},
  volume    = {538},
  doi       = {10.1051/0004-6361/201118212},
  publisher = {EDP Sciences},
}

@Article{Mannheim1993,
       author = {{Mannheim}, K.},
        title = "{The proton blazar.}",
      journal = {\aap},
         year = 1993,
        month = mar,
       volume = {269},
        pages = {67-76},
          doi = {10.48550/arXiv.astro-ph/9302006},
archivePrefix = {arXiv},
       eprint = {astro-ph/9302006},
 primaryClass = {astro-ph},
       adsurl = {https://ui.adsabs.harvard.edu/abs/1993A&A...269...67M}
}

@Article{Alves2018,
  author    = {Alves, E. P. and Zrake, J. and Fiuza, F.},
  journal   = {Physical Review Letters},
  title     = {Efficient Nonthermal Particle Acceleration by the Kink Instability in Relativistic Jets},
  year      = {2018},
  issn      = {1079-7114},
  month     = dec,
  number    = {24},
  pages     = {245101},
  volume    = {121},
  doi       = {10.1103/physrevlett.121.245101},
  publisher = {American Physical Society (APS)},
}

@Article{Nalewajko2018,
  author    = {Nalewajko, Krzysztof},
  journal   = {Monthly Notices of the Royal Astronomical Society},
  title     = {Three-dimensional kinetic simulations of relativistic magnetostatic equilibria},
  year      = {2018},
  issn      = {1365-2966},
  month     = sep,
  number    = {4},
  pages     = {4342--4354},
  volume    = {481},
  doi       = {10.1093/mnras/sty2549},
  publisher = {Oxford University Press (OUP)},
}

@ARTICLE{Narayan2012,
       author = {{Narayan}, Ramesh and {Piran}, Tsvi},
        title = "{Variability in blazars: clues from PKS 2155-304}",
      journal = {\mnras},
         year = 2012,
        month = feb,
       volume = {420},
       number = {1},
        pages = {604-612},
          doi = {10.1111/j.1365-2966.2011.20069.x},
archivePrefix = {arXiv},
       eprint = {1107.5812},
 primaryClass = {astro-ph.HE},
       adsurl = {https://ui.adsabs.harvard.edu/abs/2012MNRAS.420..604N}
}

@ARTICLE{Bugli2025,
       author = {{Bugli}, M. and {Lopresti}, E.~F. and {Figueiredo}, E. and {Mignone}, A. and {Cerutti}, B. and {Mattia}, G. and {Del Zanna}, L. and {Bodo}, G. and {Berta}, V.},
        title = "{Relativistic reconnection with effective resistivity: I. Dynamics and reconnection rate}",
      journal = {\aap},
         year = 2025,
        month = jan,
       volume = {693},
          eid = {A233},
        pages = {A233},
          doi = {10.1051/0004-6361/202452277},
archivePrefix = {arXiv},
       eprint = {2410.20924},
 primaryClass = {astro-ph.HE},
       adsurl = {https://ui.adsabs.harvard.edu/abs/2025A&A...693A.233B}
}
\bibliographystyle{aasjournal}

\end{document}